\documentclass[aps,prd,twocolumn,groupedaddress,nofootinbib]{revtex4-1}

\usepackage{graphicx}

\usepackage{rotating}

\usepackage{dcolumn}

\usepackage{amssymb}

\usepackage{bm}

\usepackage{wrapfig}

\usepackage{subfig}

\usepackage{MyPack}

\begin{document}

\title[Gravitational radiation from radial infall into a Schwarzschild black
hole]{Gravitational radiation from radial infall of a particle into a
Schwarzschild black hole. A numerical study of the spectra, quasi-normal modes
and power-law tails.}

\author{Ermis Mitsou}
\email{Ermis.Mitsou@unige.ch}
\affiliation{D\'epartement de Physique Th\'eorique, Universit\'e de 
	     Gen\`eve, CH-1211 Geneva, Switzerland}

\begin{abstract}
The computation of the gravitational radiation emitted by a particle falling
into a \Sch black hole is a classic problem studied already in the 1970s. Here we present a detailed numerical analysis of the case of radial infall starting at infinity with no initial velocity. We compute the radiated waveforms, spectra and energies for multipoles up to $l = 6$, improving significantly on the numerical accuracy of existing results. This is done by integrating the Zerilli equation in the frequency domain using the
Green's function method. The resulting wave exhibits a ``ring-down" phase
whose dominant contribution is a superposition of the quasi-normal modes of the black hole. The numerical accuracy allows us to recover the frequencies of these modes through a fit of that part of the wave. Comparing with direct computations of the quasi-normal modes we reach a $\sim 10^{-4}$ to $\sim 10^{-2}$ accuracy for the first two overtones of each multipole. Our numerical accuracy also allows us to display the power-law
tail that the wave develops after the ring-down has been exponentially cut-off.
The amplitude of this contribution is $\sim 10^2$ to $\sim 10^3$ times smaller
than the typical scale of the wave. 
\end{abstract}

\maketitle

\section{Introduction} 

The gravitational radiation due to a point-like particle falling into
a black hole (BH) is a classic problem in General Relativity whose first computation goes back to the 1970s \cite{DRPP, DRT}. Being an elementary BH perturbation, it served as one of the first numerical applications of the BH perturbation theory initially developed in the classic papers \cite{RW,Z}. The present literature in this field is abundant\footnote{See \cite{NR} for a summary of the theory and related references in theoretical and numerical studies. The textbook \cite{MGW} also contains discussions and references on the subject.} and the infalling particle model sometimes appears as a limit case of more general scenarios such as the infall of an arbitrary number of particles \cite{BC1} or, most importantly, the coalescence of BH binaries \cite{BC3}, where it corresponds to the extreme mass-ratio limit. 

The general features of the radiated waveform are by now well understood \cite{A}. The first part of the signal the observer receives is called the \textit{precursor} and
corresponds to the radiation emitted directly from the infalling source to the
observer. It is therefore insensitive to what happens near the BH and it was shown that it can be well described by a resummation of the Post-Newtonian expansion \cite{DN1,DN2,DN3,BC4}\footnote{This was actually shown in the more general case of the coalescence of a BH binary where the corresponding part of the signal is the oscillation due to the inspiral phase of the BHs.}. Then comes the \textit{ring-down} phase which is dominated by a superposition of the quasi-normal modes (QNMs) of the BH. These are characteristic information of the \Sch metric and are therefore brought by waves that were reflected in the neighborhood of the maximum of the effective potential of the problem (the ``barrier"), situated at $\sim 1.5$ \Sch radii. The QNMs are a very interesting feature also because they provide a possible bridge between classical and quantum gravity \cite{Bek,H,Ber,MQNM}. Finally, the wave exhibits a power-law tail at large values of the retarded time, after the ring-down has been exponentially cut-off. This residual radiation corresponds to waves that were not initially heading towards the observer but ended up reaching him by scattering off the background metric at large distance of the horizon, hence the delay and the decrease in amplitude.

In this paper, we focus on the simplest case, the radial infall starting at infinity with no initial velocity, as in the original works \cite{DRPP,DRT} (computations for more general initial data include for example \cite{LP,MP,BC3}). The spectrum of the radiated signal can be computed from a numerical integration of a single wave-equation, the Zerilli equation \cite{Z}. The first part of our work consists in performing this computation for multipoles up to $l=6$ with high accuracy, that is $\sim 10^{-5}$. This allows us to analyze the spectra at quantitatively higher orders than present estimates. The waveforms, which are obtained by Fourier transforming the spectra, reach an accuracy of $\sim
10^{-4}$ and $\sim 10^{-2}$ at worst. For $l \leq 4$, we are able to keep that precision on a relatively large interval of retarded time which includes the beginning of the power-law phase. Finally, we extract the QNM frequencies by fitting the ring-down phase with damped sinuses and then compare the results with the values obtained through direct computations \cite{L,K,ECS}. This way we determine how ``visible" the QNMs are. 

The organization of the paper is as follows. In section \ref{sec:th} we recall
the main formulas describing the production of gravitational radiation from a
radially in-falling test mass. In section \ref{sec:re} we present and discuss
our results. Finally, an account of computational details and error estimation is presented in the appendix.

\section{Theoretical background}\label{sec:th}

In what follows, $m$ and $M$ are the masses of the particle and BH, respectively. We use units $G = c = 1$ and the following definition of the Fourier
transform
\begin{equation} \label{eq:Fou} 
\Psi_l(r,\omega) \equiv \frac{1}{\sqrt{2\pi}} \int{\psi_l(r,t) e^{i\omega t}dt} \, .
\end{equation}
The clear distinction in mass scales between the BH and the particle allows the problem to be treated within BH perturbation theory to first order (linearized Einstein equations). Therefore, the particle is a test mass, i.e. moving along the geodesics of the unperturbed \Sch space-time, and produces a perturbation $h_{\mu\nu}$ (of the \Sch metric) which is a test tensor field (on the \Sch geometry) since we only keep first order terms of it in the Einstein equations. These can be reduced to two independent one-dimensional scalar wave-equations for each $l$ multipole (in the tensor spherical harmonics decomposition), where $l$ represents the \textit{total} angular momentum number. The wave-equation's effective potential term has no $m$ dependence since the background metric is spherically symmetric. In the case of the radial in-fall of a particle, nor does the source term because of cylindrical symmetry. Thus the $m \neq 0$ modes are not excited. At each $l$, these equations describe the dynamics of the two eigenstates of the parity operator, which don't mix because of the symmetry of the background metric under parity, and together fully determine the $2^l$-pole of $h_{\mu\nu}$. In the case of the radial infall of a particle, only the so called \textit{polar} modes\footnote{They are the ones picking $(-1)^l$ under parity, i.e. the ``true" scalar, vector, etc, as opposed to the ``pseudo" ones picking $(-1)^{l+1}$, called \textit{axial} modes.} are excited and we are left with one equation, the Zerilli equation. Denoting its solution by $\psi_l$, the 
radial dependence of the $2^l$-pole of
$h_{\mu\nu}$   is $\sim (1/r)\psi_l(t-r)$ in the radiation zone. The Zerilli equation on the frequency domain is \cite{Z}
\begin{equation} \label{eq:IDE}
\partial_{r_*}^2 \Psi_l + (\omega^2 - V_l(r)) \Psi_l = S_l(r,\omega)
\end{equation}
where as usual $r_* = r + 2M\ln(r/2M - 1)$, and the effective potential is
\begin{eqnarray}
\lefteqn{V_l(r) = \left(1 - \frac{2M}{r}\right)} \nonumber \\
& &  \hspace{6mm} \frac{2\lambda^2(\lambda+1) + 3\frac{2M}{r}\lambda^2 +
\frac{9}{2}\left(\frac{2M}{r}\right)^2\lambda +
\frac{9}{4}\left(\frac{2M}{r}\right)^3} {(\lambda r + 3M)^2} \label{eq:V}
\end{eqnarray}
with $\lambda = (l-1)(l+2)/2$. The source term is
\begin{eqnarray} 
\lefteqn{S_l(r,\omega) = -\frac{4m}{\lambda r+3M} \sqrt{l+1/2} \left(1 -
\frac{2M}{r}\right)} \nonumber \\
& &  \hspace{15mm} \left[ \sqrt{\frac{r}{2M}} - i\frac{2\lambda}{\omega(\lambda
r+3M)} \right] e^{i\omega T(r)}  \label{eq:S}\, ,
\end{eqnarray}
where $T(r)$ is determined by the geodesic of the particle in the \Sch metric,
\begin{eqnarray}
\lefteqn{\frac{T(r)}{2M} = -\frac{2}{3}\left(\frac{r}{2M}\right)^{3/2} -
2\left(\frac{r}{2M}\right)^{1/2}} \nonumber \\
& & \hspace{6mm} + \log \left[ \left(\sqrt{r/2M}+1\right)
\left(\sqrt{r/2M}-1\right)^{-1} \right] \, .\label{eq:T}
\end{eqnarray}
The differential equation (\ref{eq:IDE}) is solved with boundary conditions of
purely ingoing waves at the \Sch radius and purely outgoing ones at infinity
\begin{eqnarray} 
\lim_{r_* \to -\infty} \Psi_l(r_*,\omega) & = & A_{l,in}(\omega)e^{-i\omega r_*}
\, , \label{eq:PsiICin} \\
\lim_{r_* \to \infty} \Psi_l(r_*,\omega) & = & A_{l,out}(\omega)e^{i\omega r_*}
\, . \label{eq:PsiICout}
\end{eqnarray}
These correspond to the fact that the source is always localized in space and therefore GWs can only be emitted \textit{towards} the infinities. In order to compute the radiated amplitude of the $\omega$-mode $A_{l,out}(\omega)$, it is convenient to use the Green's function method. Let
$y_l^-(r_*,\omega)$ and $y_l^+(r_*,\omega)$ be the solutions of the homogeneous
equation of (\ref{eq:IDE}) 
\begin{equation} \label{eq:HDE}
\partial_{r_*}^2 y^{\pm}_l + (\omega^2 - V_l(r)) y^{\pm}_l = 0
\end{equation}
with boundary conditions
\begin{eqnarray} 
\lim_{r_* \to -\infty} y_l^-(r_*,\omega) & = & e^{-i\omega r_*} \, ,
\label{eq:YIC} \\
\lim_{r_* \to \infty} y_l^+(r_*,\omega) & = & e^{i\omega r_*} \, ,
\end{eqnarray}
so that they match (\ref{eq:PsiICin}) and (\ref{eq:PsiICout}) in the final
solution. At $r_* \to \infty$ the potential vanishes and $y_l^-$ tends towards the
analytical form. 
\begin{equation} \label{eq:YLL}
\lim_{r_* \to \infty} y_l^-(r_*,\omega) = \alpha_l(\omega)e^{i\omega r_*} +
\beta_l(\omega)e^{-i\omega r_*}\, .
\end{equation}
Thus the reflection and transmission coefficients of
$V_l$ for a monochromatic wave coming from plus infinity are
\begin{equation} \label{eq:RT}
R_l = \left| \frac{\alpha_l(\omega)}{\beta_l(\omega)} \right|^2 \hspace{0.8cm} T_l = \frac{1}{|\beta_l(\omega)|^2} \, ,
\end{equation}
which we also compute for reasons made clear in section
\ref{sec:ab}. The Wronskian may be written $W_l(\omega) = 2i\omega\beta_l(\omega)$ and, in the radiation zone, one obtains
\begin{eqnarray} \label{eq:Wout}
\lefteqn{\Psi_{l,out}(r,\omega) = \lim_{r \to \infty} \Psi_l(r,\omega)}
\nonumber \\
& & \hspace{8mm} = \frac{e^{i\omega r_*}}{2i\omega\beta_l(\omega)}
\int_{-\infty}^{\infty}
S_l(\tilde{r_*},\omega)y_l^-(\tilde{r_*},\omega)d\tilde{r_*}
\end{eqnarray}
which by definition of the outgoing $\omega$-mode (\ref{eq:PsiICout}) gives
\begin{equation} \label{eq:A}
A_{l,out}(\omega) = \frac{1}{2i\omega\beta_l(\omega)} \int_{-\infty}^{\infty}
S_l(\tilde{r_*},\omega)y_l^-(\tilde{r_*},\omega)d\tilde{r_*}\, .
\end{equation}
Given the choice of normalization in (\ref{eq:Fou}), the radiated energy
spectrum of the $l$-mode is \cite{Z,DRPP} 
\begin{equation} \label{eq:dEdw}
\frac{dE_l}{d\omega} = \frac{1}{32\pi} \frac{(l+2)!}{(l-2)!} \omega^2
|A_{l,out}(\omega)|^2 \, .
\end{equation}
The waveform is found using the inverse Fourier transform. Introducing the retarded time $u \equiv t - r^*$, we have
\begin{eqnarray} \label{eq:wout}
\lefteqn{\psi_{l,out}(u) = \frac{1}{\sqrt{2\pi}} \int_{-\infty}^{\infty}
\Psi_{l,out}(r,\omega)e^{-i\omega t}d\omega} \nonumber \\
& & \hspace{10mm} = \sqrt{\frac{2}{\pi}} \int_{0}^{\infty}
\Re[A_{l,out}(\omega)e^{-i\omega u}]d\omega \, .
\end{eqnarray}
The last equality comes from (\ref{eq:PsiICout}) and the fact that $\psi_l$ is
real, i.e. $A_{l,out}(-\omega) = \bar{A}_{l,out}(\omega)$. So the procedure
consists in computing $y_l^-$ through eq.~(\ref{eq:HDE}) with initial condition
given by eq.~(\ref{eq:YIC}), and then evaluating (\ref{eq:A}) by extracting $\beta_l(\omega)$ out of
eq.~(\ref{eq:YLL}) and performing the integral. Once we have $A_{l,out}(\omega)$,
the energy spectrum is given by (\ref{eq:dEdw}) and the wave-function is
computed through eq.~(\ref{eq:wout}).

\section{Results} \label{sec:re}

In this section we present the results of our numerical integration. A detailed account of the numerical procedure and error estimation is given in appendices~\ref{ch:comp} and \ref{ch:err}, respectively. From now on we simplify the
notation by dropping the ``out" subscript in $A_{l,out}(\omega)$ and
$\psi_{l,out}(\omega)$ and writing $f_l(\omega) \equiv dE_l/d\omega$. 
The figures are collected at the end of the paper.

\subsection{Analysis of the frequency spectrum} \label{sec:resAf}

The top and middle panels of figure \ref{fig:All_l} give the energy spectrum $f_l(\omega)$ in both linear and logarithmic scales, the modulus $|A_l(\omega)|$ and phase $\phi_l(\omega)$ (of $A_l(\omega))$. The shape of the energy spectra is in agreement with the result of ref.~\cite{DRPP}. In table \ref{tab:Espec} we list the radiated energies for every $2^l$-pole, that is 
\begin{equation} \label{eq:Etotl}
E_l = \int_{0}^{\infty} f_l(\omega) d\omega \, . 
\end{equation}
To better characterize
the energy spectrum, we have also computed the values $(\omega_l^{\star},f_l^{\star})$ at
which $f_l(\omega)$ is maximal and the following quantities:
\begin{eqnarray} 
\langle\omega\rangle_l & \equiv & \frac{4M^2}{E_l} \int_{0}^{\infty} \omega f_l(\omega)
d\omega \, ,  \label{eq:mw} \\
\sigma_l(\omega) & \equiv & \sqrt{\langle\omega^2\rangle_l -
{\langle\omega\rangle}_l^2} \, . \label{eq:sw}
\end{eqnarray}
We also use $f_t(\omega) \equiv \sum_{l\leq 6} f_l(\omega)$ as an estimation of the total spectrum.

\begin{table}[h] 
\centering
\begin{tabular}{|c|l|l|l|l|l|} 
  \hline
  $l$  &  $M/m^2E_l$  &  $f_l^{\star}/m^2$  &  $2M\omega_l^{\star}$  & 
$2M\langle\omega\rangle_l$  &  $2M\sigma_l (\omega)$ \\
  \hline
  2  &  $9.1368(9) 10^{-3}$  &  $3.5943(4) 10^{-2}$  &  $0.61992(9)$ 
&  $0.5224(1)$  &  $0.1961(6)$  \\
  3  &  $1.1004(1) 10^{-3}$  &  $3.3977(3) 10^{-3}$  &  $1.0534(1)$  &
 $0.8747(2)$  &  $0.271(1)$  \\
  4  &  $1.4947(1) 10^{-4}$  &  $4.0757(4) 10^{-4}$  &  $1.4685(2)$  &
 $1.226(2)$  &  $0.329(2)$ \\
  5  &  $2.1380(2) 10^{-5}$  &  $5.3971(5) 10^{-5}$  &  $1.8688(2)$  
&  $1.582(3)$  &  $0.375(3)$ \\
  6  &  $3.1423(3) 10^{-6}$  &  $7.5273(8) 10^{-6}$  &  $2.2726(2)$  &
 $1.941(4)$  &  $0.413(4)$ \\
\hline
  t  &  $1.0411(1) 10^{-2}$  &  $3.7744(4) 10^{-2}$  &  $0.6236(2)$  &
 $0.5723(1)$  &  $0.2529(1)$ \\
  \hline
\end{tabular}
\caption[Multipole total radiated energies]{From left to right: Multipole total
radiated energy, energy spectrum's maximal value, corresponding frequency, mean value and standard deviation (for $f_l(\omega)$ seen as a
distribution).}
\label{tab:Espec}
\end{table}

The estimation $E_t$ of the total radiated energy (bottom left of table \ref{tab:Espec}) is
to be compared with the value $\simeq 0.0104 m^2/M$ given in \cite{DRPP}. To a first approximation, $E_l$ seems to follow the exponential trend
proposed in \cite{DRPP}, which is $\sim e^{-2 l}$. We find that the form
\begin{equation} \label{eq:Etrend}
E_l \simeq 0.56(3) l^{0.9(2)} e^{-1.75(5) l} m^2/M 
\end{equation}
actually provides a better fit.
Assuming that energies for $l > 6$ follow this empirical law, we get the order
of the contribution of the neglected multipoles in the total energy, that is $E_{l>6} \sim 10^{-7} m^2/M$, which is less than our numerical precision for $E_t$. As for the maximum of $f_t(\omega)$, we find it is situated at $M\omega_t^{\star} = 0.3118(1)$ whereas \cite{DRPP} finds $\simeq 0.32$.

We now study the asymptotic behavior of the spectra.

\subsubsection{Low frequency limit}  

In this limit we find that our numerical results are very well fitted by
\begin{equation} \label{eq:logphi}
\log|\phi_l(\omega \ll 1) -\phi_l(\omega=0) | \simeq a^{\phi}_l + b^{\phi}_l \log(\omega)
\end{equation}
and
\begin{equation}  \label{eq:logf}
\log(f_l(\omega \ll 1)) \simeq a^f_l + b^f_l \log(\omega) \, ,
\end{equation}
that is, $\phi_l$ and $f_l$ have a power-law behavior. Within our numerical precision, we find that, for the multipoles $l=2,\ldots, 6$ that we have studied,  $\phi_l(\omega=0)$ is very well reproduced by\footnote{Not directly computable because of the $\omega^{-1}$ term in the source term, but through extrapolation.}
\begin{equation}\label{eq:ph0}
\phi_l(\omega=0)= \frac{\pi (l-3)}{6}\, .
\end{equation}
In order to obtain the coefficients $a^{\phi}_l,b^{\phi}_l,a^f_l,b^f_l$ with great accuracy
it is necessary to perform the fit in the very low frequency region. In Table~\ref{tab:LF} we show these coefficients, obtained by performing a fit at $2M\omega\sim 10^{-4}$, and we find that 
\begin{equation} \label{eq:ftrend}
f_l(\omega \ll 1) \sim \omega^{2 l/ 3} \, 
\end{equation}
reproduces our results very well. 

\begin{table}[h]
\centering
\begin{tabular}{|c||r|r||r|r|}
  \hline
  $l$  &  $a^{\phi}_l$  &  $b^{\phi}_l$  &  $a^f_l$  &  $b^f_l$   \\
  \hline
   2  &  $1.38(1)$  &  $0.706(3)$  &  $-1.84(6)$  &  $1.321(8)$  \\
  3  &  $1.54(2)$  &  $0.707(2)$  &  $-4.40(5)$  &  $1.989(7)$  \\
  4  &  $1.64(1)$  &  $0.707(1)$  &  $-6.92(4)$  &  $2.653(6)$  \\
 \hline
\end{tabular}
\caption[Low frequency behavior coefficients.]{The first few fitting coefficients
in eqs.~(\ref{eq:logphi}) and (\ref{eq:logf}).}
\label{tab:LF}
\end{table}

For $l=2$ we are even able to go down to $2M\omega\sim 10^{-6}$ and we get 
\begin{equation}\label{eq:fl2ex}
f_{l=2}(\omega \ll 1) \simeq 0.176(3)\, (2M\omega)^{1.332(2)} m^2 \, .
\end{equation}
For each given $l$, the low frequency asymptotic behavior for $f_l(\omega)$ and $\phi_l(\omega=0)$ can be computed analytically because it is due to the motion of the particle far from the horizon, where the trajectory can be well approximated by the Newtonian one and the GW emission can be computed using the multipole expansion in flat space.
For $l=2$ the contribution comes from the mass quadrupole and the computation is performed in  ref.~\cite{Wag}
(see also  section 4.3.1 of ref.~\cite{MGW}). The result is
\begin{eqnarray}
f_{l=2}(\omega\ll 1)
&=& \left(\frac{2}{3}\right)^{7/3} \frac{\Gamma^2(1/3)}{5\pi}\,\, (2M \omega)^{4/3} m^2\nonumber\\
&\simeq &0.1774\,\,  (2M\omega )^{4/3} m^2 \label{eq:fl2th}\, ,
\end{eqnarray}
so our numerical result (\ref{eq:fl2ex}) reproduces very well the exact analytic behavior. This is a significant check of our numerical procedure. We have also performed this analytical computation for $l=3$ and again got agreement with eqs.~(\ref{eq:ph0}) and (\ref{eq:ftrend}). Both equations can be combined into
\begin{equation} \label{eq:Alf}
A_l(\omega \ll 1) \simeq a_l (i\omega)^{(l-3)/3} \, ,
\end{equation}
where $a_l$ is a positive real number. Finally, the $\phi_l(\omega)$ variation, which is not computable analytically, is found to be constant with respect to $l$ within our error margins
\begin{equation} 
\phi_l(\omega \ll 1) - \phi_l(\omega = 0) \sim \omega^{0.707(5)} \, .
\end{equation}

\subsubsection{High frequency limit}

For this limit the full relativistic treatment is necessary and there is no simple analytical expression to compare with. Here we
only focus on $f_l$. The top right panel of figure \ref{fig:All_l} clearly
suggests an exponential cutoff. We find that the fitting form
\begin{equation} \label{eq:Phf}
\log f_l(\omega \gg 1) \simeq a_l + b_l \omega + c_l \log(\omega) \, ,
\end{equation} 
gives the best fit. Table \ref{tab:HF} lists the resulting values for the parameters. 

\begin{table}[h] 
\centering
\begin{tabular}{|c|r|r|r|}
  \hline
  $l$  &  $a_l$  &  $b_l$  &  $c_l$  \\
  \hline
  2  &  $5.34(2)$  &  $-12.34(4)$  &  -3.03(8)  \\
  3  &  $10.06(1)$  &  $-12.26(3)$  &  -4.20(9)  \\
  4  &  $14.81(1)$  &  $-12.18(3)$  &  -5.42(3)   \\
  5  &  $19.73(1)$  &  $-12.17(4)$  &  -6.60(8)   \\
  6  &  $24.81(2)$  &  $-12.15(5)$  &  -7.7(2)   \\
  \hline
\end{tabular}
\caption[High frequency behavior coefficients.]{Fitting coefficients in eq. (\ref{eq:Phf}).}
\label{tab:HF}
\end{table}

\subsection{Waveforms and quasi-normal modes}

In figures \ref{fig:l2} to \ref{fig:l6} we show, for $l = 2,\dots,6$, the amplitude spectrum $A_l(\omega)$, the waveform $\psi_l(u)$, the fit of its ringdown phase using QNMs and its power-law tail (compared to the obtained QNM fit). 

In the waveform plots, we observe that the number of significant oscillations in the
ring-down increases with $l$ while its typical length appears to be the same
for all $l$. The precursor's typical length decreases with $l$. The
typical amplitude decreases quite fast, as suggested by the empirical law in
the corresponding energy (\ref{eq:Etrend}). 

As for the recovery of the QNM frequencies by fitting the ring-down phase, the procedure has already worked very well in numerical studies on the coalescence of BH binaries \cite{BCP,BC4}. It is therefore expected to work well in the case of the infalling particle model since it is a special case of the latter and numerically much simpler to treat. Table \ref{tab:QNMfit} lists the results of the fit for the first three overtones while table \ref{tab:QNMex} lists the values obtained through direct numerical computations \cite{L,K,ECS} (see appendix \ref{ch:comp}, eq. (\ref{eq:QNMt}) for the fitting form). On the top panel of figure \ref{fig:QNM} we see that our error margins cover the expected values only for the first overtone $(n = 1)$. Looking at the plots with the fitting curves, we see that the first mode's domination interval is pretty much the same for all $l$. On the
other hand, the power-law contribution sets in later or is relatively weaker
with increasing $l$. The high-$l$ graphics are indeed the ones were the QNMs
fitted the wave best, so the QNM contribution becomes more ``visible" with increasing $l$. Inversely, in the case of $l = 2$, the power-law tail contribution sets
in so early that it prevents us from even having a visible superposition of fit
and data. 

Finally, concerning the tail of the wave, analytical studies \cite{A} show that it is actually a superposition of power-laws and that one has to go quite far in retarded time in order to see the leading term dominate. However, at such high values our precision breaks down so we cannot perform any useful fit but we still display the graphic result for $l = 2,3,4$ at lower $u$. The error on the $l = 5,6$ cases is already notable at relatively low $u$. 

\begin{table}[h]
\begin{tabular}{|c|r|r|r|}
  \hline
  $l$  &  $n=1$  &  $n=2$  &  $n=3$  \\
  \hline
  2  &  $0.747(9) - i0.17(2)$  &  $0.45(3) - i0.51(9)$  &    \\
  3  &  $1.198(2) - i0.18(1)$  &  $1.104(8) - i0.54(1)$  &  $0.93(4) - i1.5(3)$ \\
  4  &  $1.616(1) - i0.186(6)$  &  $1.580(8) - i0.552(6)$  &  $1.34(3) -
i1.140(4)$  \\
  5  &  $2.025(1) - i0.190(4)$  &  $1.97(1) - i0.556(9)$  & $1.82(4) -
i0.895(8)$  \\
  6  &  $2.424(1) - i0.190(4)$  &  $2.378(8) - i0.572(9)$  &  $2.22(4) -
i0.805(4)$  \\
  \hline
\end{tabular}
\caption{Computed QNM frequencies through waveform's fit.}
\label{tab:QNMfit}
\end{table}

\begin{table}[h]
\begin{tabular}{|c|r|r|r|}
  \hline
  $l$  &  $n=1$  &  $n=2$  &  $n=3$  \\
  \hline
  2  &  $0.7473 - i0.1779$  &  $0.6934 - i0.5478$  &    \\
  3  &  $1.1989 - i0.1854$  &  $1.1653 - i0.5626$  &  $1.1034 - i0.9582$  \\
  4  &  $1.6184 - i0.1883$  &  $1.5933 - i0.5687$  &  $1.5454 - i0.9598$  \\
  5  &  $2.0246 - i0.1897$  &  $2.0044 - i0.5716$  &  $1.9654 - i0.9607$  \\
  6  &  $2.4240 - i0.1905$  &  $2.4071 - i0.5733$  &  $2.3741 - i0.9611$  \\
  \hline
\end{tabular}
\caption{Direct computation QNM frequencies.}
\label{tab:QNMex}
\end{table}

\begin{acknowledgments} 
I would like to thank Michele Maggiore for proposing
this work, for his continuous interest in the developments of the
study and guiding suggestions, in both ideas and literature. I also thank Vitor Cardoso and Alessandro Nagar for their useful comments, suggestions and for considerably broadening my knowledge on the present state of the research in this field. Finally, I am grateful to Kostas Kokkotas for providing me with the exact QNM values for $l = 4,5$ and $6$ and to Andreas Malaspinas for helping me with the computer resources.
\end{acknowledgments}

\appendix

\section{Computational details} \label{ch:comp}

In this section we present all the algorithms and techniques involved in the calculations. From now on we use only dimensionless variables. Thus, in what follows, $\omega$, $r_*$, $A_{l,out}$, $f_l$, $E_l$ and $\psi_{l,out}$ actually stand for $2M\omega$, $r_*/2M$, $A_{l,out}/(Mm)$, $f_l/m^2$, $M/m^2 E_l$ and $\psi_{l,out}/m$, respectively. Relative errors and margins are denoted using the symbol $\delta$ whereas $\Delta$ is used to denote absolute errors or integration grid steps, depending on the context.

\subsection{The $\beta_l(\omega)$ parameter} \label{sec:ab}

Consider equation (\ref{eq:YLL}). The convergence towards that asymptotic form being too slow, we use the next order terms (as in \cite{LP}):
\begin{eqnarray} \label{eq:MYLL}
y_l^-(r_* \gg 1,\omega) \simeq \alpha_l(\omega)e^{i\omega r_*} + \beta_l(\omega)e^{-i\omega r_*} + \nonumber\\
\frac{1}{\omega r_*} \left( \gamma_l(\omega)e^{i\omega r_*} + \delta_l(\omega)e^{-i\omega r_*} \right)
\end{eqnarray}
We take 47 equally spaced sample points over 5 typical periods\footnote{These values are chosen so as to increase the information input in our fit and also avoid repeating values which lead to a badly conditioned system.} $2\pi/\omega$ on $y_l^-$ and plug them in (\ref{eq:MYLL}). Having four complex unknowns, this gives us an overdetermined linear system $A x = b$, where $A$ is a $47 \times 4$ complex matrix and $x = (\alpha_l, \beta_l, \gamma_l, \delta_l)$. Then multiplying by $A^T$ on the left we get a $4 \times 4$ matrix equation
\begin{equation} 
A^T A x = A^T b \, ,
\end{equation}
the solution of which minimizes the residue $||A x - b||_2^2$. This equation is then solved using the Gaussian elimination method. 

This routine is used repeatedly at increasing $r_*$, along with the computation of the integral in (\ref{eq:A}), thus making the result more and more precise\footnote{This is actually true only up to a certain limit because $A^T A$ contains three different orders of magnitude: ($\sim 1$), ($\sim 1/r_*$) and ($\sim 1/r_*^2$), so if $r_*$ is too large the problem is badly conditioned. In our case, however, we didn't reach that point.}. Note that there is no direct check on the error of $\beta_l(\omega)$. However, the fluctuations of its value strongly affecting $A_l(\omega)$, we take the latter's good convergence as a guarantee for an acceptable error on $\beta_l(\omega)$. A bit of monitoring at various stages of the computation confirms that the convergence of the $\beta_l(\omega)$ value is a lot faster than that of the integral in (\ref{eq:A}). Another source of confidence is the value $R_l(\omega)+T_l(\omega)$ (see eq. (\ref{eq:RT})) whose fluctuation around $1$ also gives qualitative information on that error. In practice, we find a standard deviation of $\sim 10^{-6}$ on the $\omega$-grid.

\subsection{The radiation amplitude $A_l(\omega)$} \label{sec:compA}

For simplicity we write $A$ for $A_l(\omega)$ since we describe the computation at a given value of $\omega$ and $l$. We use Numerov's method to integrate $y_l^-$ in (\ref{eq:HDE}). Starting at finite $r_* \ll 0$, the initial condition corresponding to (\ref{eq:YIC}) is actually $e^{-i\omega r_*} \sum_{k>0} a_k (r-1)^k$ for some constants $a_k = O(1)$, at least for small $k$. We start at $r_* = -700$ because that's where $r-1$ starts being computable ($\sim 10^{-300}$), in double precision. This allows us to neglect the correction to the initial condition (\ref{eq:YIC}). Since the source term is also negligible at $r_* = - O(100)$, the integral in (\ref{eq:A}), which is computed in the same loop as $y_l^-$, does not need any corrections for stating at finite $r_*$ either. We use the trapezoidal rule for this integration because comparison with other Newton-Cotes formulae shows that it is the fastest method, in the sense that it starts approximating well at already big grid steps $\Delta r_*$. This is due to the oscillatory nature of the integrand making positive and negative errors approximately compensate each other. The overall convergence is dramatically slow because the source term (\ref{eq:S}) goes asymptotically as $\sim r^{-1/2}$. But the more we continue the more the integration's error grows. Therefore, we have to use a special method in order to extract that limit faster.

We define $A(r_*)$ as being the value of (\ref{eq:A}) where the integral is truncated on the upper bound at $r_*$, so that $A = \lim_{r_* \to \infty} A(r_*)$ (remember, the $\omega$ and $l$ dependence is implicit here). On the actual $r_{*k}$ grid (the discretized axis), the computed sequence $A(r_{*k})$ approximating that value gives a damped oscillation. We divide the $r_*$-axis into intervals of length $L$ which is given by a few typical periods $2\pi/\omega$ and index them by $n = 0,1,2,\dots$. We choose to consider only the average value of $A(r_{*k})$ out of every such interval, call it $A_n$ (its phase $\phi_n$), a new sequence sharing the same limit. So $L$ is a smoothing parameter. We also calculate $\beta_l$ (and $\alpha_l$) at every $n$ using the procedure described in section \ref{sec:ab}. The routine stops when the last ten values of $|A_n|^2$ and $\phi_n$ are all within an $\epsilon = 10^{-4}$ margin around their respective mean values and gives the latter as a result for $A$. The $\epsilon$ margin is actually a relative one for $|A|^2$ whereas it is an absolute one for $\phi$. Thus, if we let the greatest distances from the mean values be denoted by $\Delta(|A|^2)$ and $\Delta\phi$, the condition reads $\delta(|A|^2) \equiv \Delta(|A|^2)/|A|^2 < \epsilon$ and $\Delta\phi < \epsilon$. 

Once that loop has finished, we start all over again but with half the previous $\Delta r_*$ step. This goes on until the difference between two consecutive such computations is again less than $\epsilon$ (again, relative for $|A|^2$ and absolute for $\phi$). When finished, we pass to the next $\omega$ value.

As for the $\omega$-grid parameters, let $\Delta\omega_l$ be the step and $\omega_{l,max}$ be the maximum value for which we perform the previous computation. It appears that $\omega_{l,max} \sim (l+1)$ is a good choice, as can be seen by looking at the $A_l(\omega)$ plots (top left panels of figures \ref{fig:l2} to \ref{fig:l6}) where we have set $(l+1)/2$ for the maximum of the displayed $\omega$ axis. This is why we chose $\Delta\omega_l = 6(l+1)10^{-5}$, so that the precision is the same for all $l$. However, we did not choose the $\omega_{l,max}$ value to follow the $\sim (l+1)$ trend. We find instead that there is a natural limit on the $\omega$ axis for the convergence of the $A_l(\omega)$ computation, given our precision criteria. After a given value for $f_l$, apparently common to all $l$ the program keeps dividing the $\Delta r_*$ step without ever meeting the required precision. Since the values at $\omega \gg 1$ are important for the computation of the tail of the waveform, we set the $\omega_{l,max}$ value the higher we can, that is to that natural limit. This also sets $l = 6$ as our limit for $l$, because the computation for $l = 7$ would not give enough points for the spectrum at large frequencies.

\subsection{The radiated waveform $\psi_l(u)$ and energy spectrum integrals $E_l, \langle \omega \rangle_l$ and $\sigma_l(\omega)$} \label{sec:comppsi}

We consider equations (\ref{eq:wout}), (\ref{eq:Etotl}), (\ref{eq:mw}) and (\ref{eq:sw}). We use a high order Newton-Cotes formula for their integration, the one with 7 stages and of order 8, named Weddle's formula. We also use Richardson's extrapolation on the integrals computed with steps $\Delta\omega$ and $2\Delta\omega$ in order to further increase our precision.

For $\psi_l(u)$, the $l = 2$ case must be treated carefully because $A_{l=2}(\omega)$ diverges at the origin. It is true that, for any value of $l$, we can only compute $A_l(\omega)$ for $\omega > 0$ because of the $\omega^{-1}$ factor in the source term. However, we know from (\ref{eq:Alf}) that $A_{l=3}(\omega = 0)$ is finite and therefore deducible by extrapolation and $A_{l>3}(\omega = 0) = 0$. For $l=2$, the $[0,\Delta\omega]$ contribution cannot be extrapolated. It can however be approximated through the integrand's analytic behavior at $\omega \to 0$ (discussed in section \ref{sec:resAf}, neglecting the $\phi(\omega)$ variation). Plugging the later in $\int_{0}^{\Delta\omega} \Re[A_{l=2}(\omega)e^{-i\omega u}]d\omega$ we obtain a solution involving generalized hypergeometric functions and we Taylor expand them until the desired precision is reached.

\subsection{Extraction of the QNMs out of the signal} \label{sec:compQNM}

We know that the QNMs are getting more damped with increasing $n$. Thus, if we look far enough in the ringdown phase of the wave, the least damped mode ($n = 1$) should dominate. However, the more we go at large $u$ to look for the first mode the more the tail contribution becomes notable. So our fitting model is 
\begin{equation} \label{eq:QNMt}
B_{l,n} e^{\omega_{l,n,\Im} u} \sin (\omega_{l,n,\Re} u + \theta_{l,n}) + h_{l,n} \, ,
\end{equation}
where the last term is an offset which can help compensate the shift due to the slowly appearing tail. It is really necessary for low $l$ but becomes negligible for high $l$. Once we have found the first overtone ($n = 1$), we subtract it from the waveform and reapply the fit seeking the second one ($n = 2$) and so on. 

Since there is no prescription for finding the ideal interval on the $u$ axis to perform the fit we run a small program which tries all possible intervals in $[0,30]$, retaining every time the value of the resulting fitting parameters. This gives us a 3D plot for each one of them, where the floor axes are given by the values of the interval's boundaries. The plots for $\omega_\Re$ and $\omega_\Im$ exhibit some flat areas at equal height corresponding to the ensemble of intervals at which the fit was optimal. We then cut the plot in horizontal slices of a given thickness $\Delta z$ and create a histogram giving the number of points which lie inside each one of those slices. The maximum of that histogram gives us the researched value with absolute error $\Delta z/2$ but when the spike is not clear enough we take half its width for an estimation of the error instead. For $n > 1$ we sometimes obtain more than one maximum so we choose based on the coherence with the rest of the data and by looking at the 3D plot in order to identify ``false" flat areas. The bottom of figure \ref{fig:QNM} gives an example of the 3D graphic and its histogram. As for the $B$ and $\theta$ parameters in (\ref{eq:QNMt}), no such flat area is obtained after the first fit series so we run the program one more time but with the $\omega_\Re$ and $\omega_\Im$ values already inserted.

\section{Error estimation} \label{ch:err}

\subsection{Radiated spectrum $A_l(\omega)$ and $f_l(\omega)$} \label{sec:erA}

Almost all the computed points meet the precision criteria of section \ref{sec:compA}, meaning an estimated $\sim 10^{-5}$ precision on $|A_l|^2$ and $\phi$ and therefore on $|A_l|$ and $f_l$ (remember that it is an absolute error for $\phi$). The only exceptions are for $l \geq 4$ where points with $\omega$ close to zero ($\omega \sim 10^{-3}$) have an error of $\sim 10^{-3}$. However, $|A_l|$ is very small there and the absolute error it causes in the calculation of $E_l$ and $\psi_l$ is thus negligible. In order to decrease it as much as possible anyway, we have extrapolated those regions using the analytical low frequency behavior discussed in section \ref{sec:resAf}.

As for the asymptotic behavior fits, in most cases such as in the high frequency region, there is no prescription for the ideal interval to fit on, so we estimate the error on the fitting parameters by their variation when fitting on different intervals. If, however, the number of points is small, leaving no choice about the fitting interval, the error corresponds to a 95\% c.l.

\subsection{Radiated energy spectrum characteristics (table~\ref{tab:Espec})}

There are three types of error in the computation of the integrals $E_l, \langle \omega \rangle_l$ and $\sigma_l(\omega)$: the one due to the relative error on the integrand, noted $\delta_i$, the one due to the discreteness of the integration domain, noted $\delta_d$, and the one due to its finiteness, noted $\delta_f$. $\delta_d$ is estimated by the relative difference of two integrations with grid steps $\Delta\omega$ and $2\Delta\omega$ and $\delta_f$ is estimated using (\ref{eq:Phf}). For $E_l$ we have $\delta_i \sim 10^{-5}$ since $f_l(\omega) \sim |A_l(\omega)|^2$, $\delta_d \sim 10^{-7}$ and $\delta_f \sim 10^{-8}$ so $\delta(E_l) \sim 10^{-5}$. For $\langle \omega \rangle_l$ and $\sigma_l(\omega)$, we use
\begin{equation}
\Delta p(x_1,...,x_n) \approx \sum_{k=1}^{n} \left| \frac{\partial p}{\partial x_k} \right| \Delta x_k 
\end{equation}
for the $\delta_i$ and find $\delta_i (\langle \omega \rangle_l) = 2.10^{-4}$ and $\delta_i (\sigma_l(\omega)) = 2.10^{-4}(1+2\frac{\langle \omega \rangle_l}{\sigma_l(\omega)})$. The $\delta_d$ and $\delta_f$ are again smaller. Finally, for the maximum $f_l^{\star}$ we record the smallest gap between its value and its direct neighbor's on the $\omega$ grid. However, this value is inside $f_l^{\star}(1\pm\epsilon)$, so the relative error for $f_l^{\star}$ is also given by $\epsilon$. For the error on $\omega_l^{\star}$ we simply take half the $\omega$ grid step.

\subsection{Radiated waveform $\psi_l(u)$}

First of all, being interested in the domain $u \in [-50,200]$ for $l \leq 4$ and $[-50,50]$ for $l = 5$ and 6, the greatest gap between two consecutive $\omega u$ values in the oscillatory term of (\ref{eq:wout}) is $(u\Delta\omega)_{max} = 0.06 \ll \pi$. Thus the $\omega$-grid is dense enough to take into account even the sharper variations of the integrand. The sources of error are the same as in the previous section although here we are going to use the absolute analogues for $\delta_f$ and $\delta_d$. The relative ones vary a lot near the zeros of the waveform and aren't therefore very meaningful. $\Delta_f$ is estimated using (\ref{eq:Phf}) and (\ref{eq:dEdw}) to obtain the behavior of $|A_l(\omega)|$
\begin{equation}
|A_l(\omega \gg 1)| \approx \sqrt{128\pi \frac{(l-2)!}{(l+2)!}} e^{(a_l + b_l\omega)/2} \omega^{c_l/2-1}  \, .
\end{equation}
Then
\begin{equation}
\Delta_{f,l}(u) \approx \frac{1}{\sqrt{2\pi}} \int_{\omega_{l,max}}^\infty |A_l(\omega \gg 1)| \cos(u \omega) d\omega
\end{equation}
is a good estimation of the absolute error due to the neglected part of the frequency domain. Table \ref{tab:erpsi} shows the typical values for $\Delta_d$ and $\Delta_f$ which fluctuate well inside the given order. It also gives the maximum values of the corresponding waveforms in order to compare the scales. 

\begin{table}[h] 
\centering
\begin{tabular}{|c||r|r||r|}
  \hline
  $l$  &  $\Delta_d$  &  $\Delta_f$  &  $|\psi|_{l,\max}$ \\
  \hline
  2  &  $10^{-8}$  &  $10^{-5}$  &  $4.2\times 10^{-1}$   \\
  3  &  $10^{-8}$  &  $10^{-9}$  &  $3.8\times 10^{-2}$  \\
  4  &  $10^{-9}$  &  $10^{-8}$  &  $6.3\times 10^{-3}$   \\
  5  &  $10^{-9}$  &  $10^{-10}$  &  $1.2\times 10^{-3}$  \\
  6  &  $10^{-10}$  &  $10^{-11}$  &  $2.6\times 10^{-4}$  \\
  \hline
\end{tabular}
\caption{Typical values of $\Delta_d$ and $\Delta_f$ and reference scales}
\label{tab:erpsi}
\end{table}

The regions where the relative error is maximal are of course the ones near zeros and towards the end of the tail ($\sim 10^{-2}$ at worst for the latter). Otherwise, the majority of points has $\sim 10^{-4}$ or even $10^{-5}$. Finally, the error due to the one of the integrand is $\delta_i \sim 10^{-5}$ (see section \ref{sec:compA}). 

As for the QNM, we have already explained how the errors are computed (see section \ref{sec:compQNM}).

\onecolumngrid

\begin{center}

\begin{figure}
\begin{tabular}{cc}
$\frac{1}{m^2}\frac{dE_l}{d\omega}$  vs. $2M\omega$  &  $\frac{1}{m^2}\frac{dE_l}{d\omega}$ (logarithmic scale) vs. $2M\omega$  \\
\includegraphics[width=9cm]{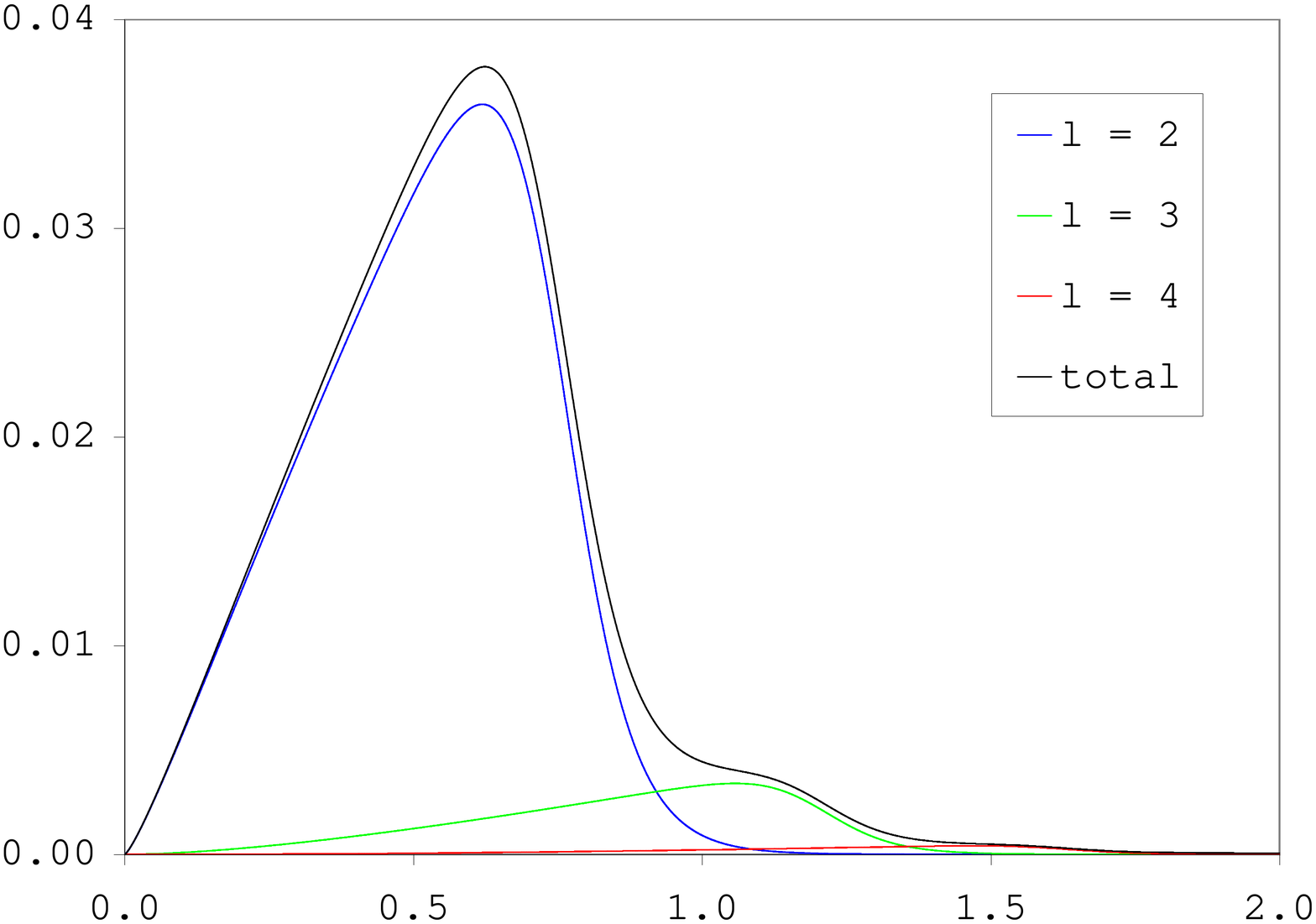}  &  \includegraphics[width=9cm]{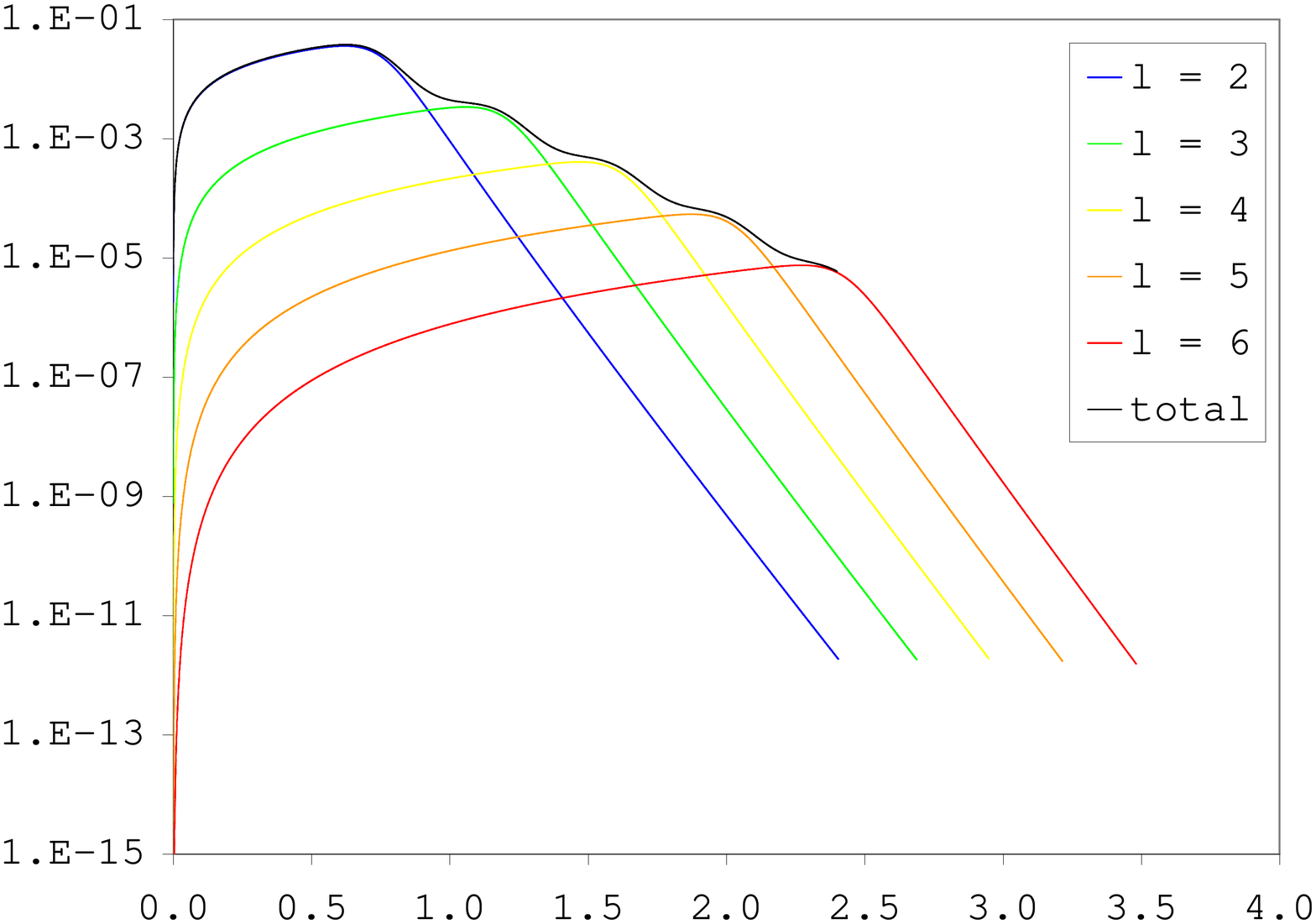} \\
  \\
$|A_{l,out}|/Mm$ vs. $2M\omega$  &  $\phi_l$ vs. $2M\omega$  \\
\includegraphics[width=9cm]{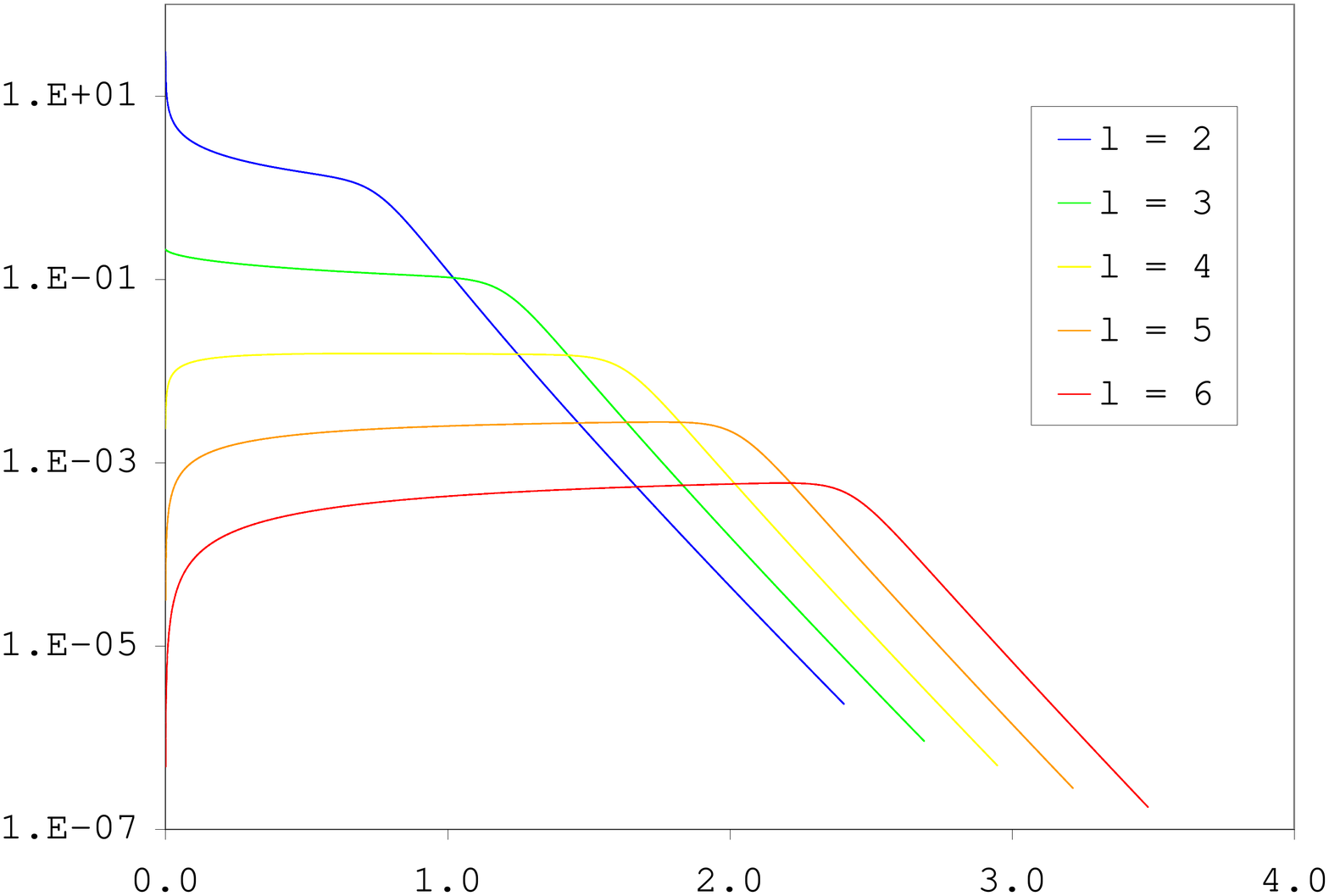}  &  \includegraphics[width=9cm]{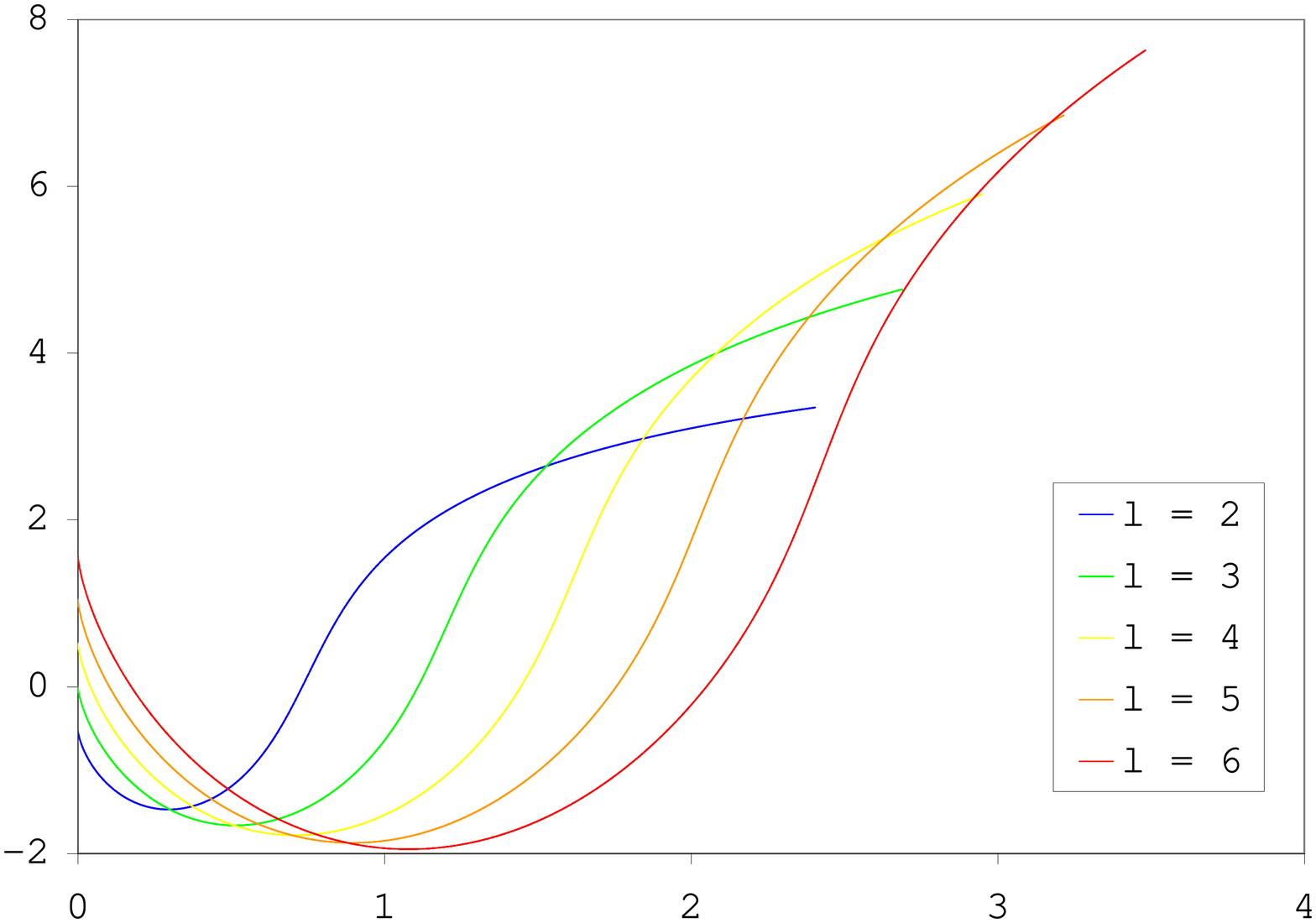} \\
 \\
$\log|\phi_l - \pi (l-3)/6|$ vs. $\log(2M\omega)$  &  $\log \left( \frac{1}{m^2}\frac{dE_l}{d\omega} \right)$ vs. $\log(2M\omega)$  \\
\includegraphics[width=9cm]{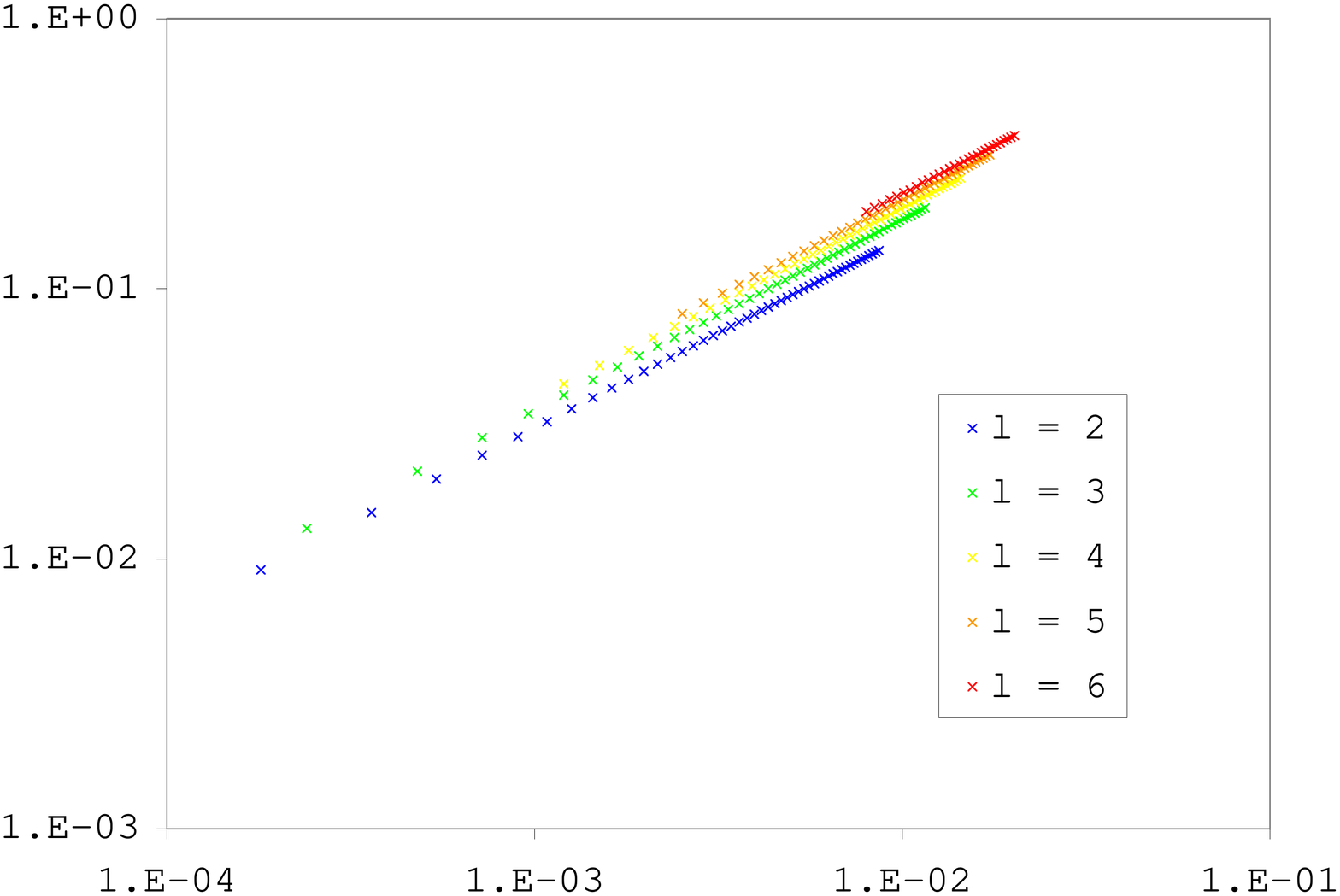}  &  \includegraphics[width=9cm]{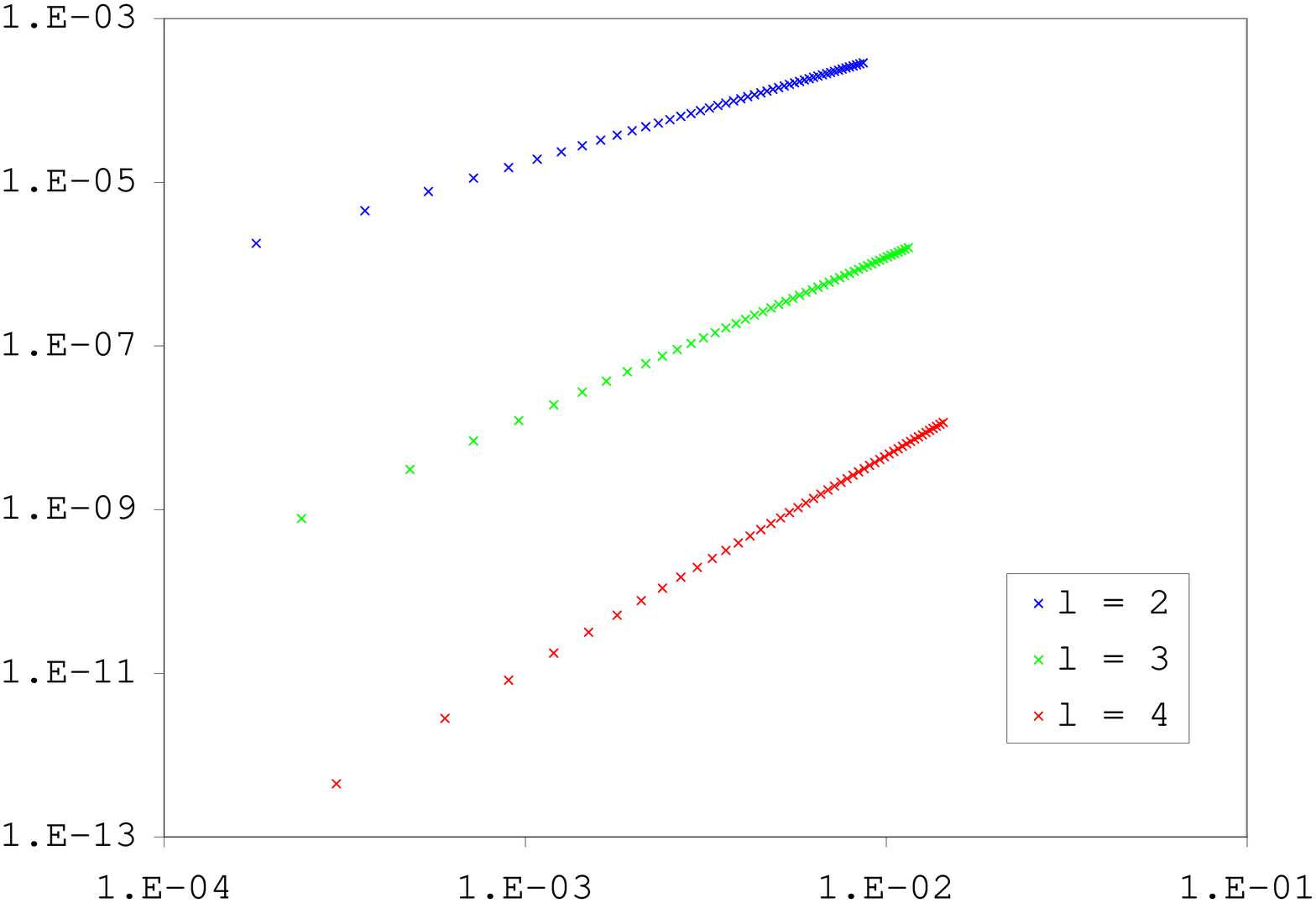} \\
\end{tabular}
\caption[Amplitude and energy spectra.]{Top panels: the energy spectra $1/m^2
dE_l/d\omega$ on normal and logarithmic scale. Middle panels: The amplitude spectra, modulus (logarithmic scale) and phase. Bottom panels: low frequency asymptotic behavior of the energy spectra and phases.}
\label{fig:All_l}  
\end{figure}

\begin{sidewaysfigure}
\begin{tabular}{cc}
$A_{2,out}/Mm$ vs. $2M\omega$  &  $\psi_{2,out}/m$ vs. $u/2M$   \\
\includegraphics[width=10cm]{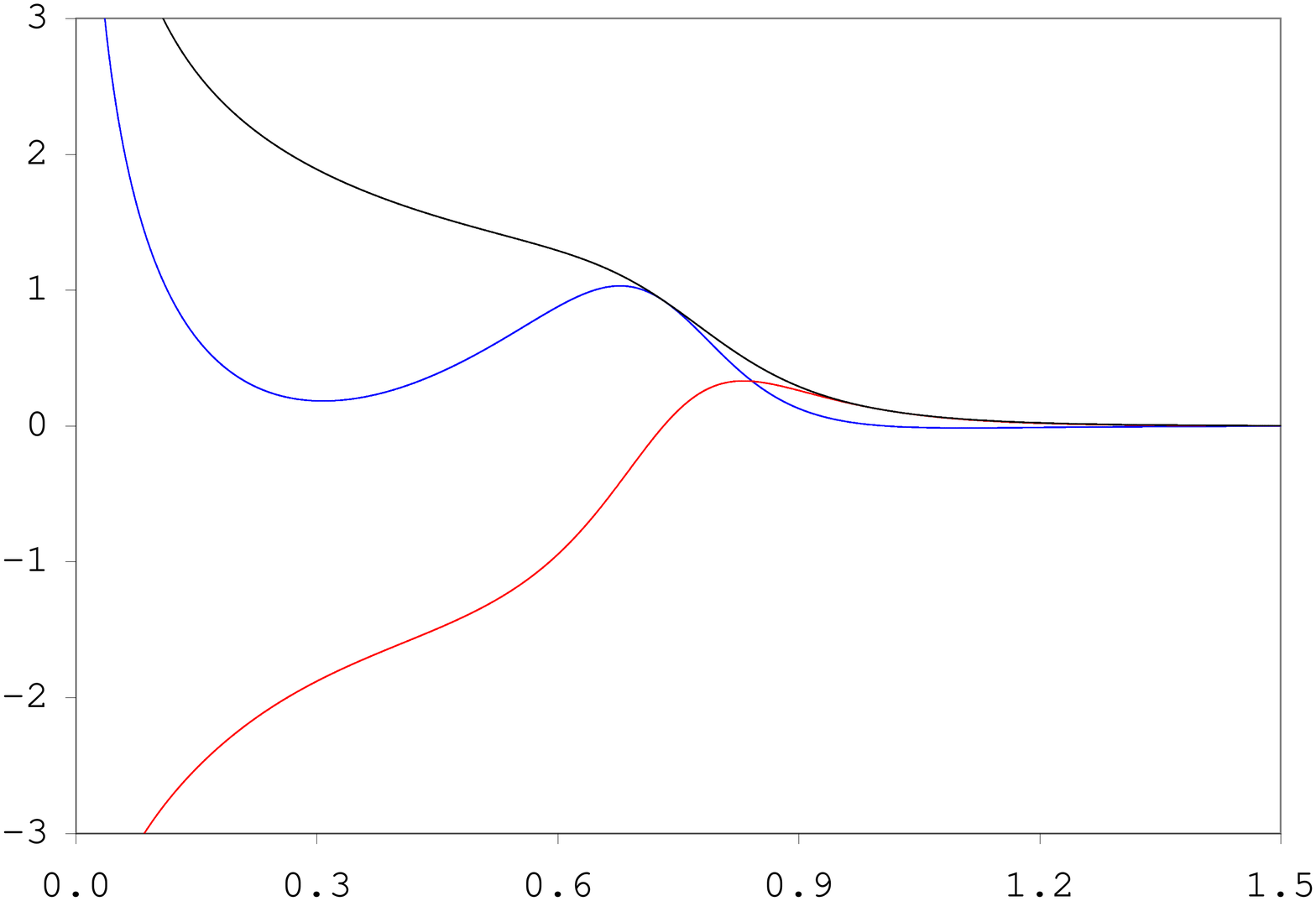}  &  \includegraphics[width=10cm]{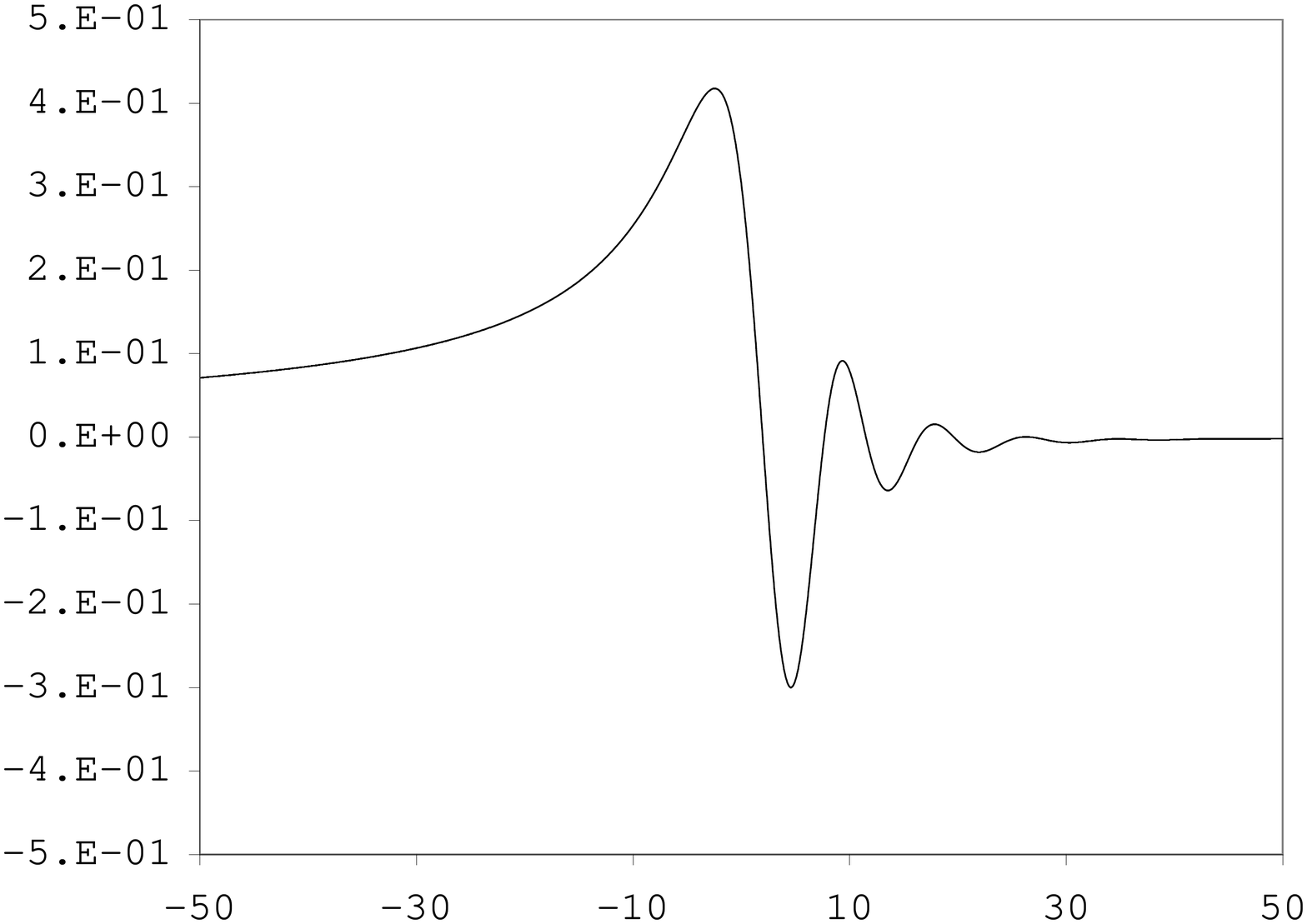}  \\
  &  \\
$\psi_{2,out}$ QNM fit  &  $\psi_{2,out}$ tail  \\
\includegraphics[width=10cm]{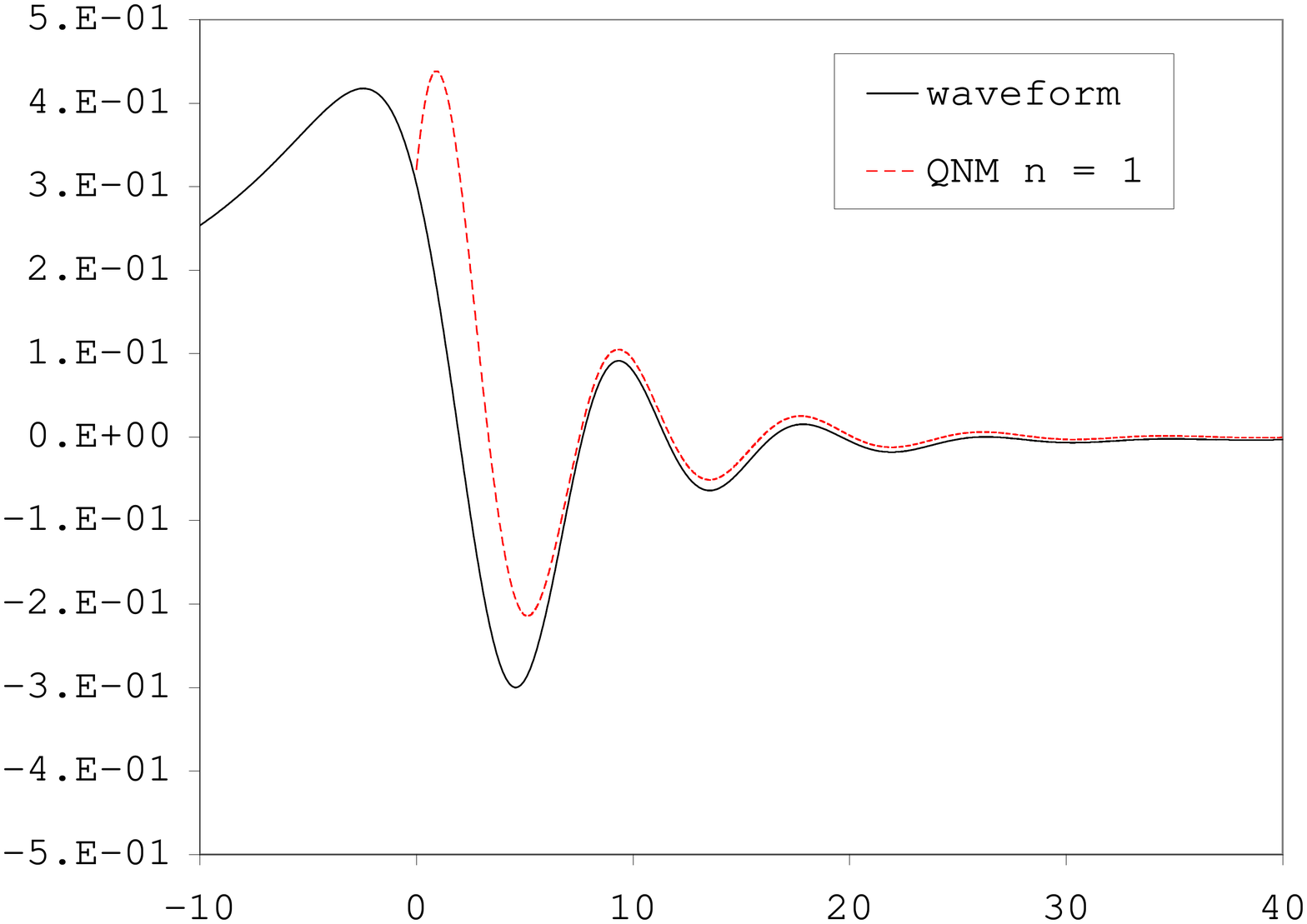}  &  \includegraphics[width=10cm]{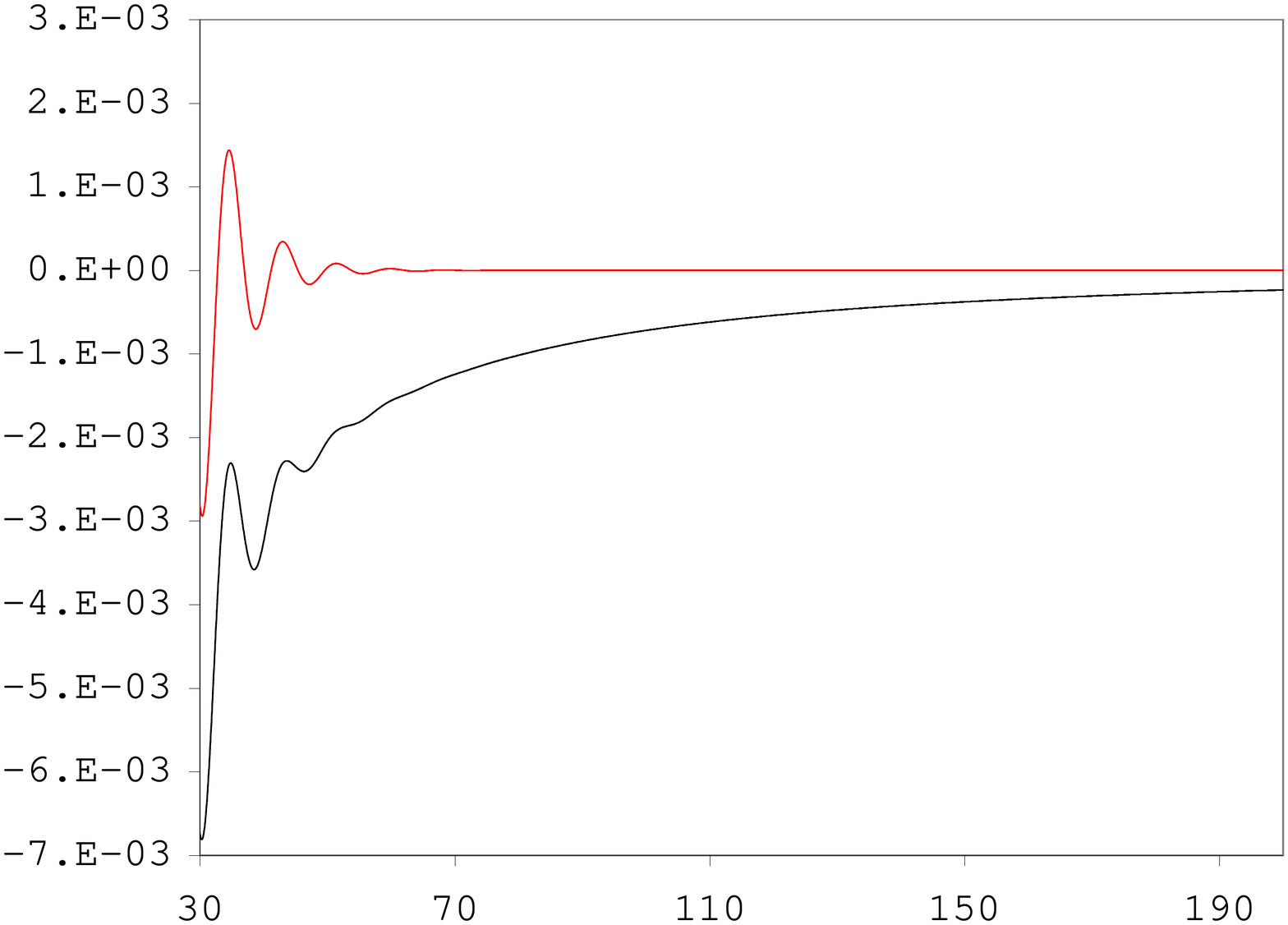}  
\end{tabular}
\caption[Spectrum, waveform, QNM fit and tail for $l = 2$]{Top left panel: the
amplitude $A_{2,out}/Mm$. Top right panel: the waveform $\psi_{2,out}/m$. Bottom left panel: the curve corresponding to the first mode of the QNM contribution. Bottom
right panel: the power-law tail of $\psi_{2,out}/m$ compared to the QNM
contribution.}
\label{fig:l2}  
\end{sidewaysfigure}

\begin{sidewaysfigure} 
\begin{tabular}{cc}
$A_{3,out}/Mm$ vs. $2M\omega$  &  $\psi_{3,out}/m$ vs. $u/2M$   \\
\includegraphics[width=10cm]{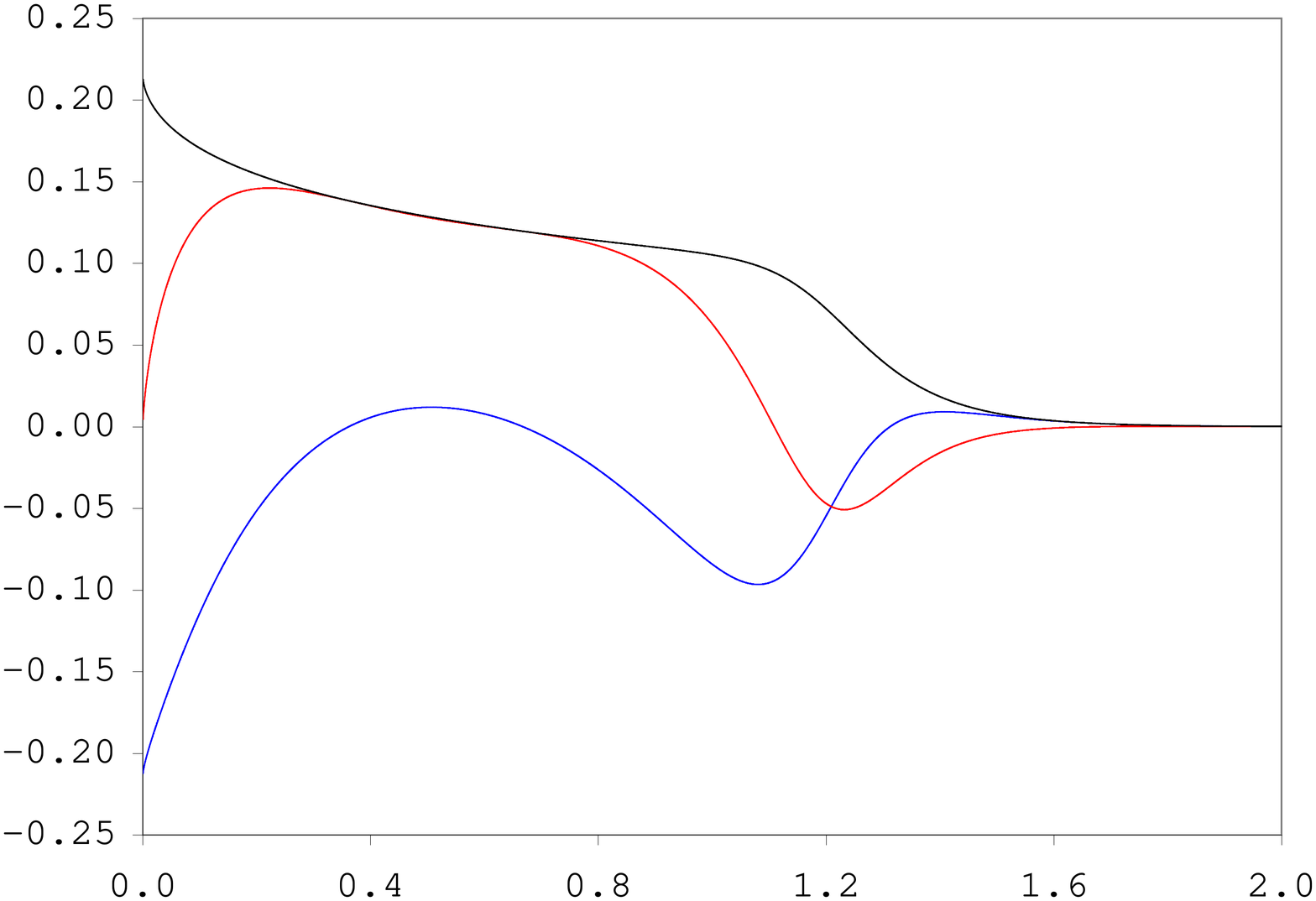}  &  \includegraphics[width=10cm]{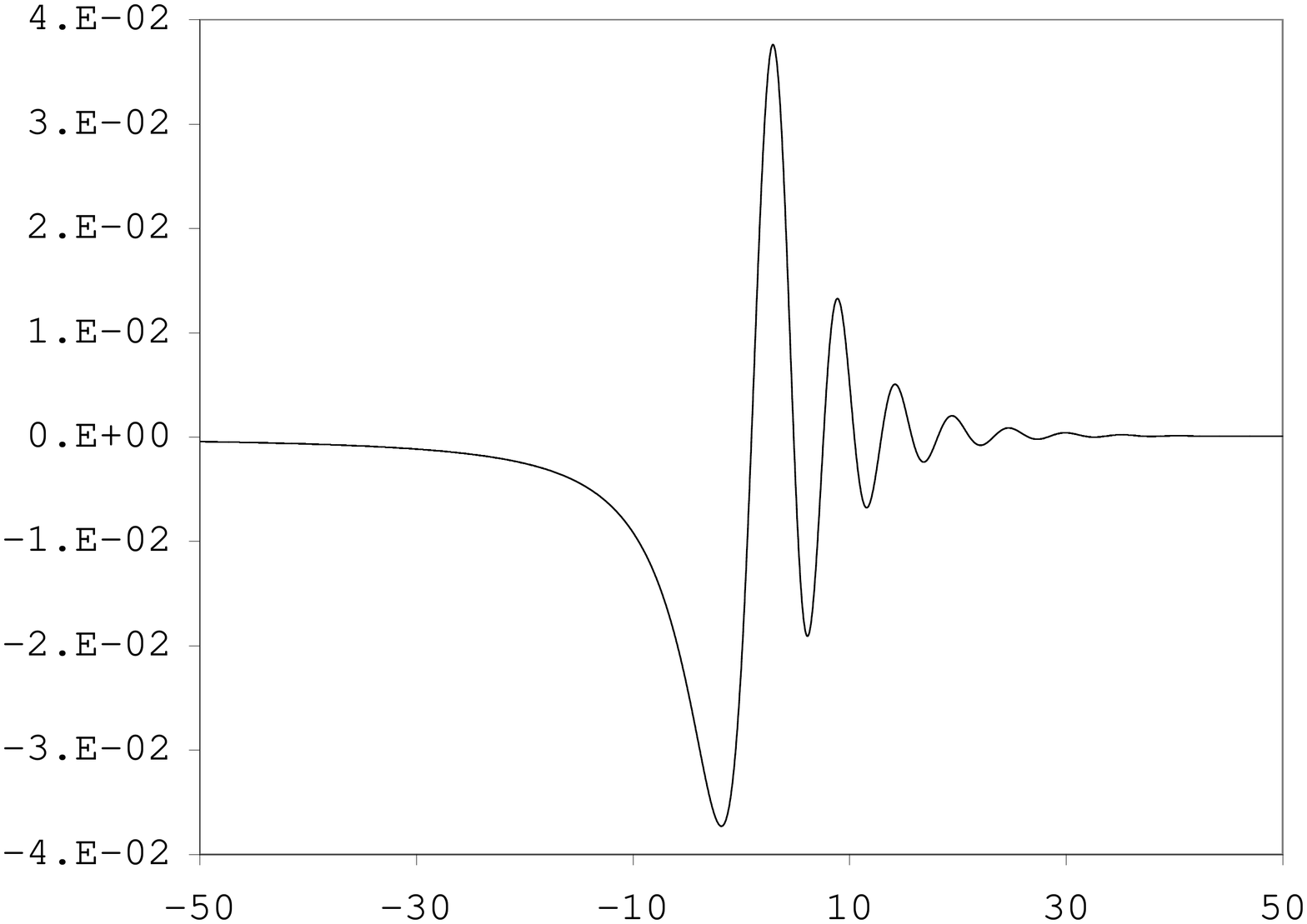}  \\
  &  \\
$\psi_{3,out}$ QNM fit  &  $\psi_{3,out}$ tail  \\
\includegraphics[width=10cm]{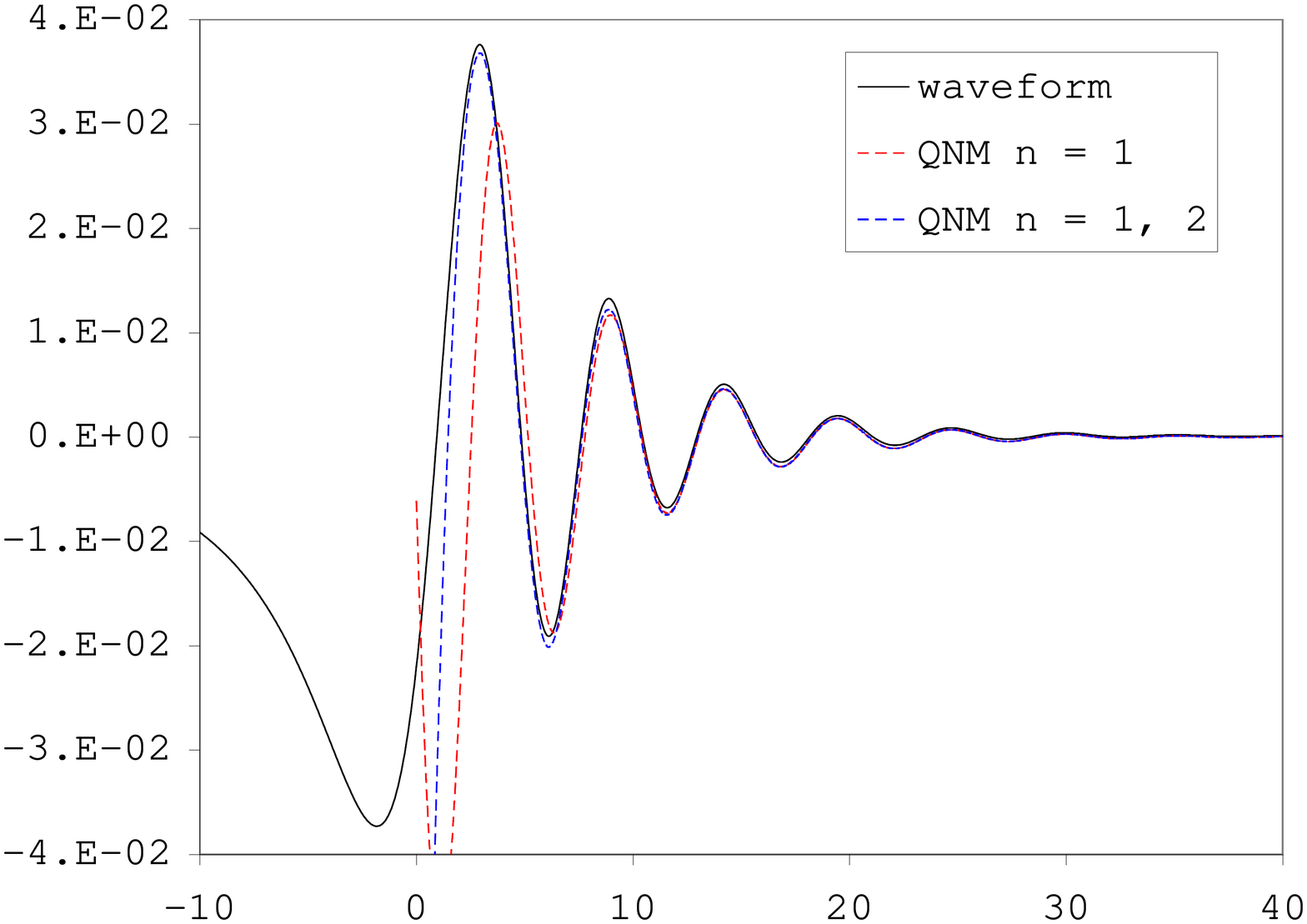}  &  \includegraphics[width=10cm]{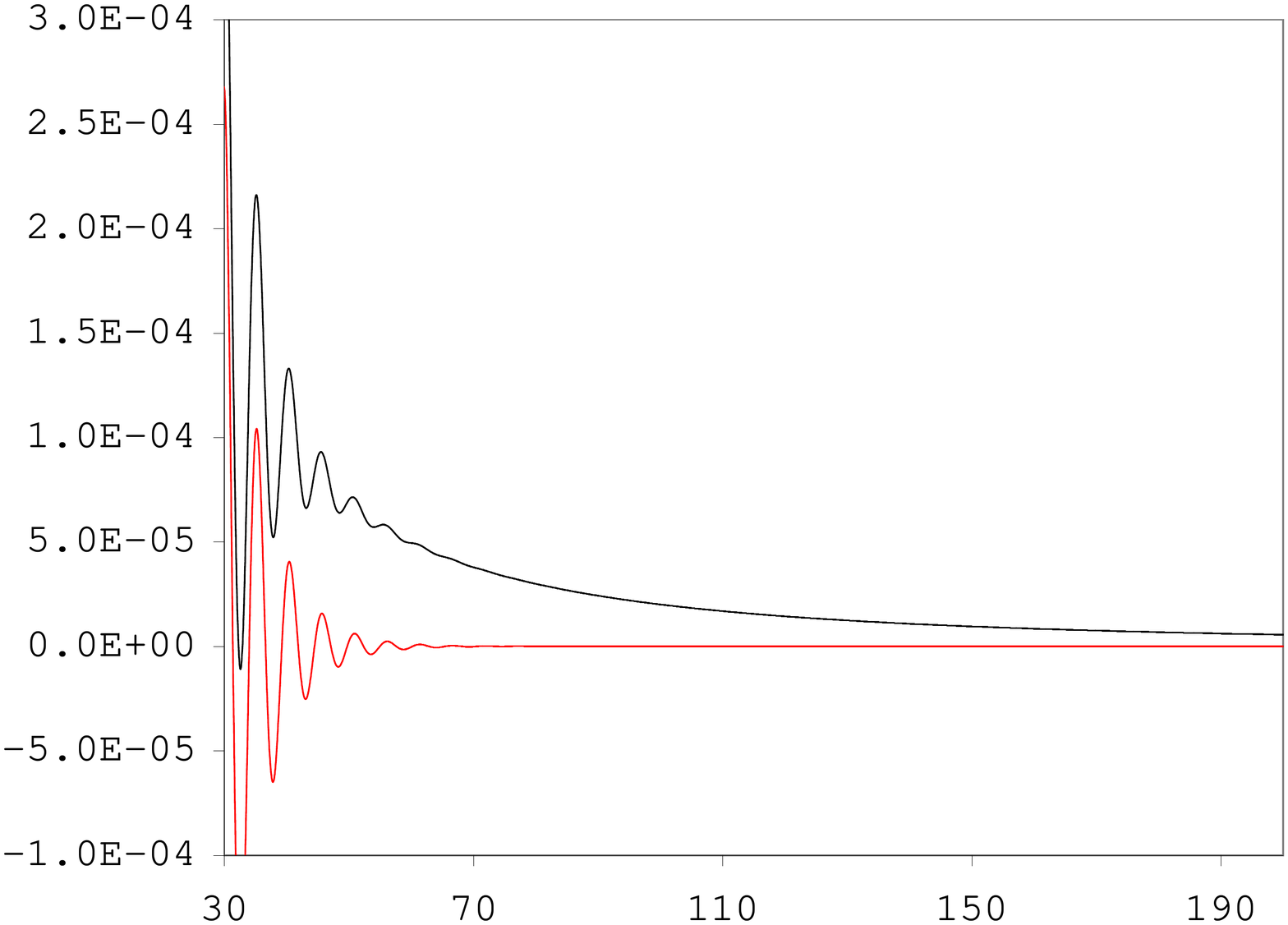}  
\end{tabular}
\caption[Spectrum, waveform, QNM fit and tail for $l = 3$]{Top left panel: the
amplitude $A_{3,out}/Mm$. Top right panel: the waveform $\psi_{3,out}/m$. Bottom left panel: the curves corresponding to the QNM contribution up to the second
overtone $n = 2$. Bottom right panel: the power-law tail of $\psi_{3,out}/m$ compared
to the QNM contribution.}
\label{fig:l3}  
\end{sidewaysfigure}

\begin{sidewaysfigure} 
\begin{tabular}{cc}
$A_{4,out}/Mm$ vs. $2M\omega$  &  $\psi_{4,out}/m$ vs. $u/2M$   \\
\includegraphics[width=10cm]{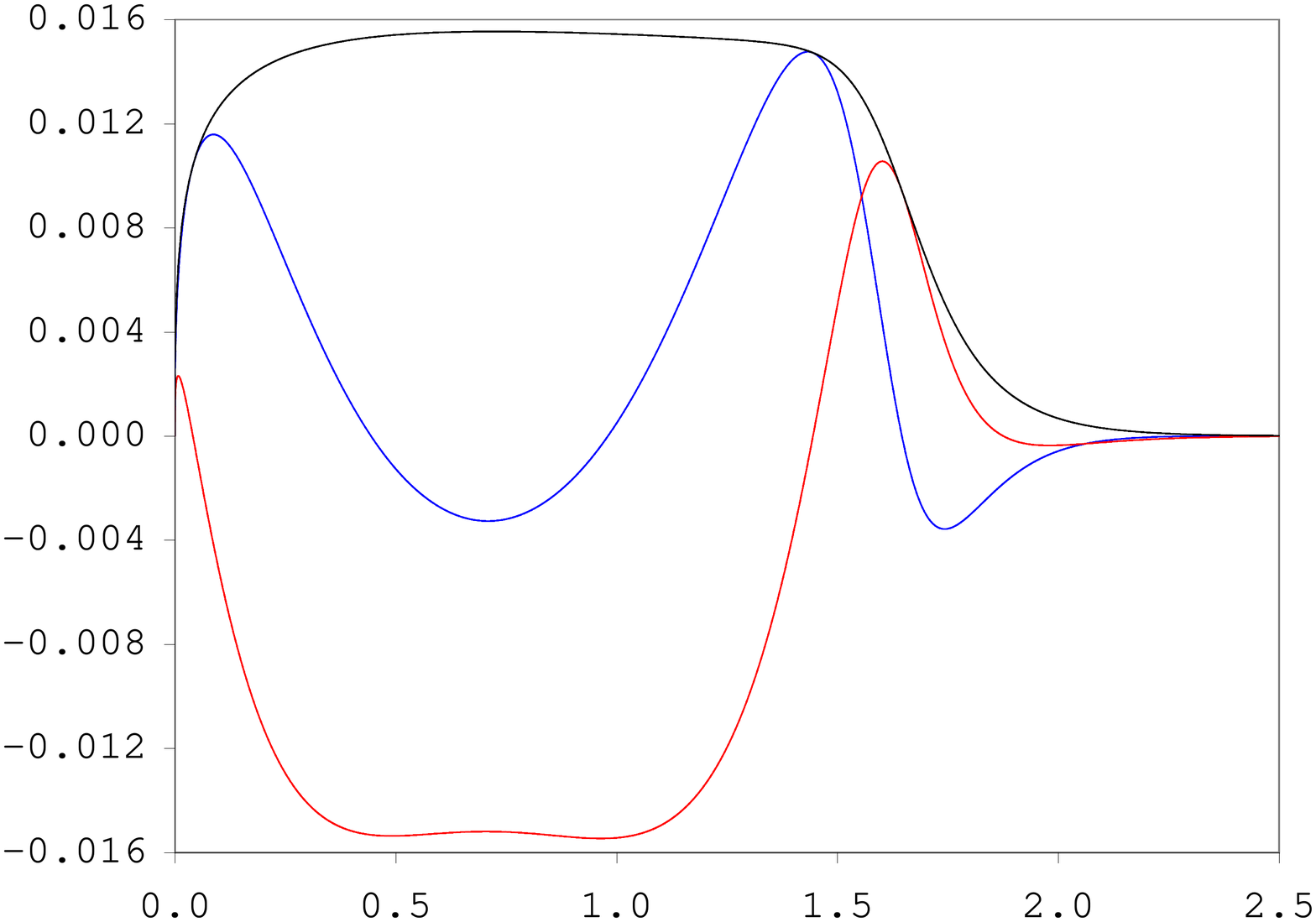}  &  \includegraphics[width=10cm]{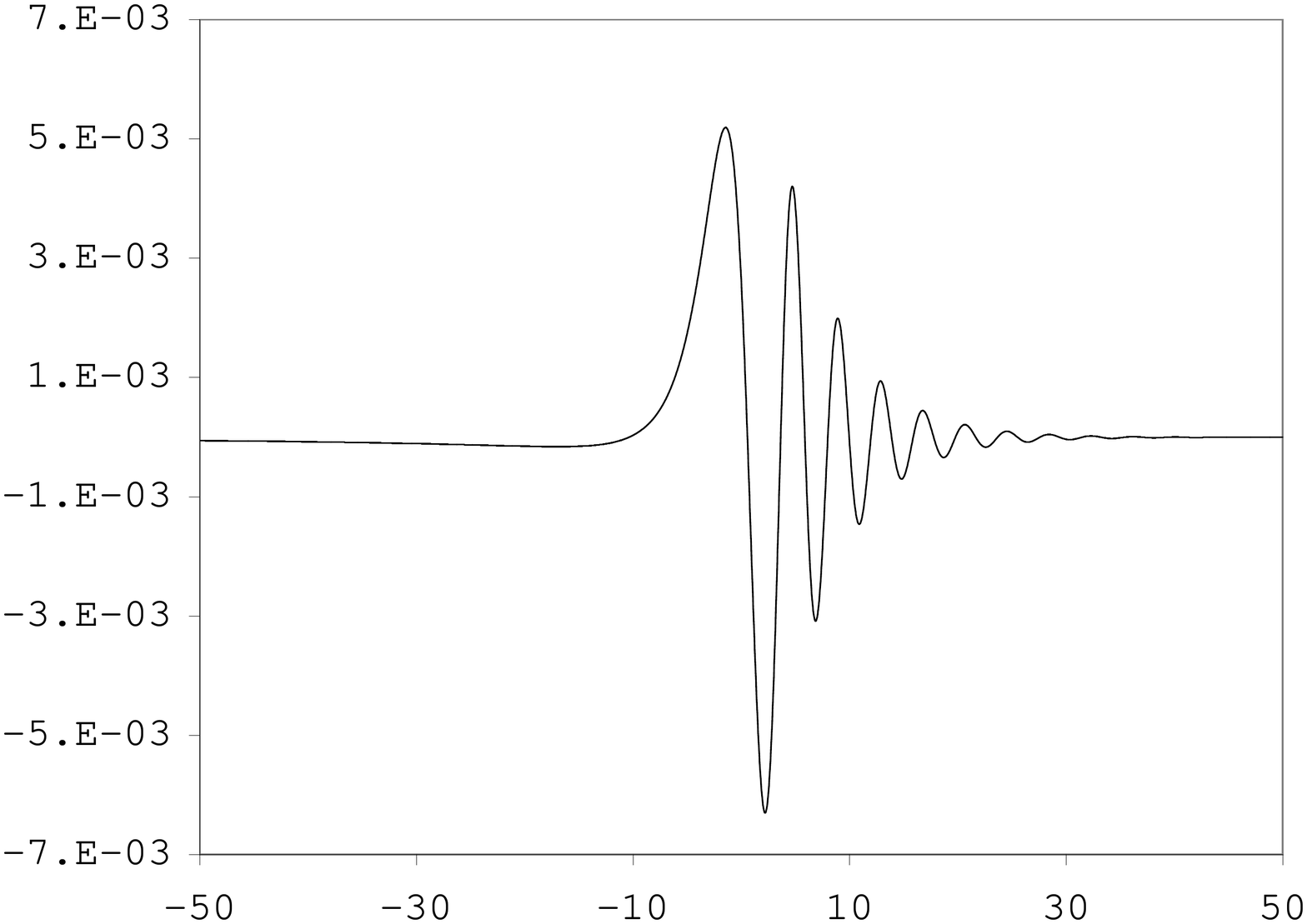}  \\
  &  \\
$\psi_{4,out}$ QNM fit  &  $\psi_{4,out}$ tail  \\
\includegraphics[width=10cm]{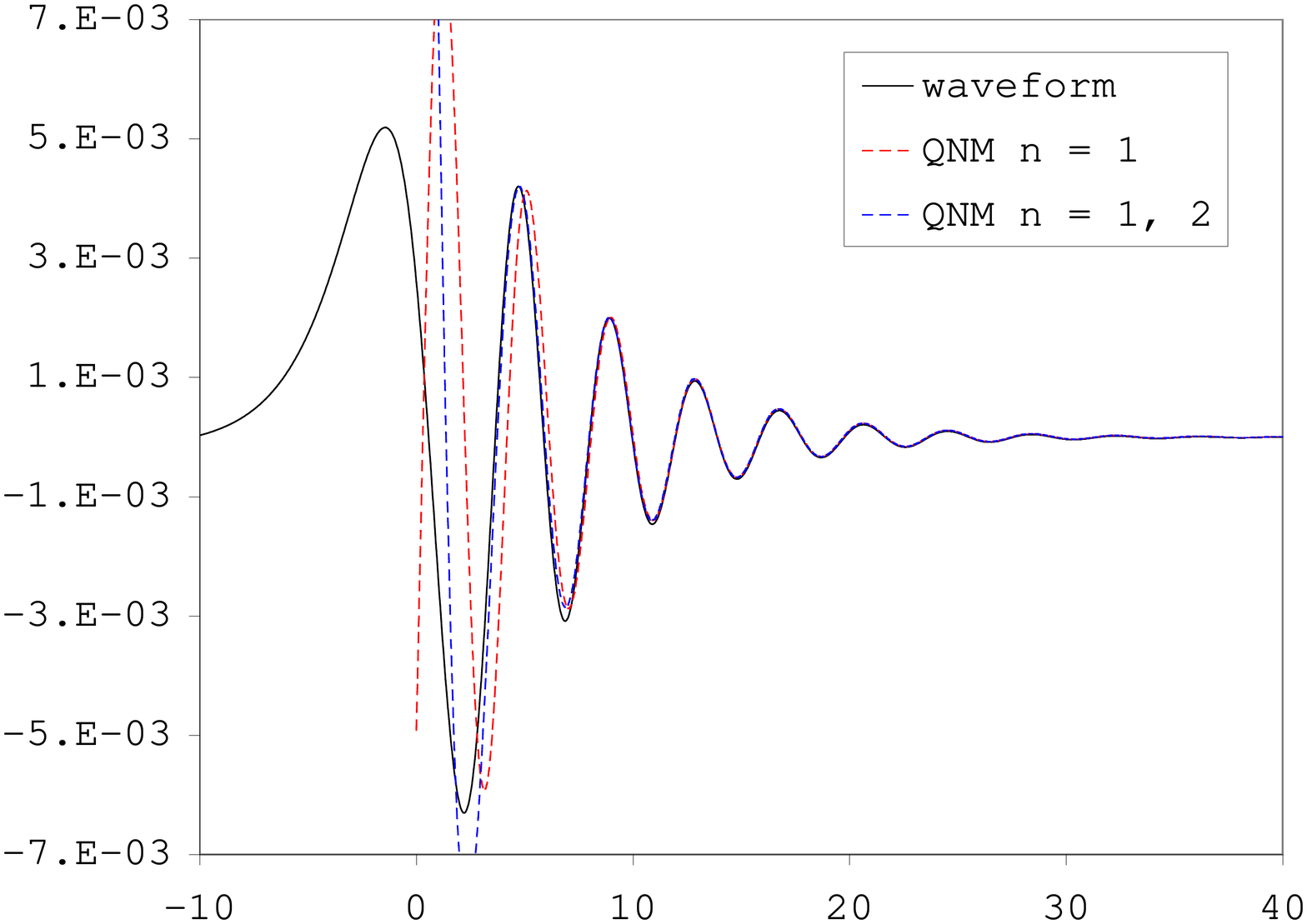}  &  \includegraphics[width=10cm]{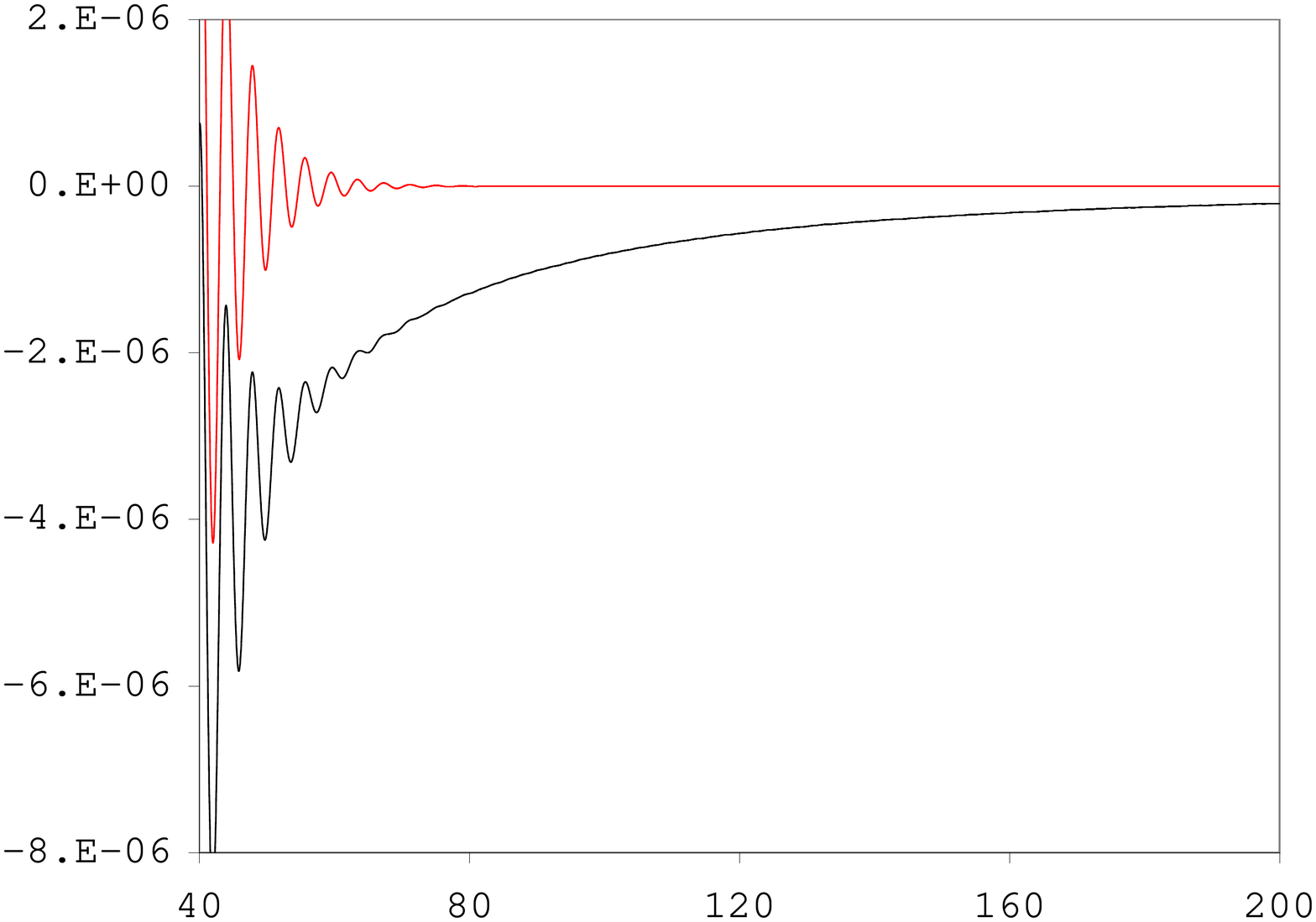}  
\end{tabular}
\caption[Spectrum, waveform, QNM fit and tail for $l = 4$]{Top left panel: the
amplitude $A_{4,out}/Mm$. Top right panel: the waveform $\psi_{4,out}/m$. Bottom left panel: the curves corresponding to the QNM contribution up to the second
overtone $n = 2$. Bottom right panel: the power-law tail of $\psi_{4,out}/m$ compared
to the QNM contribution.}
\label{fig:l4}  
\end{sidewaysfigure}

\begin{sidewaysfigure} 
\begin{tabular}{cc}
$A_{5,out}/Mm$ vs. $2M\omega$  &  $\psi_{5,out}/m$ vs. $u/2M$  \\
\includegraphics[width=10cm]{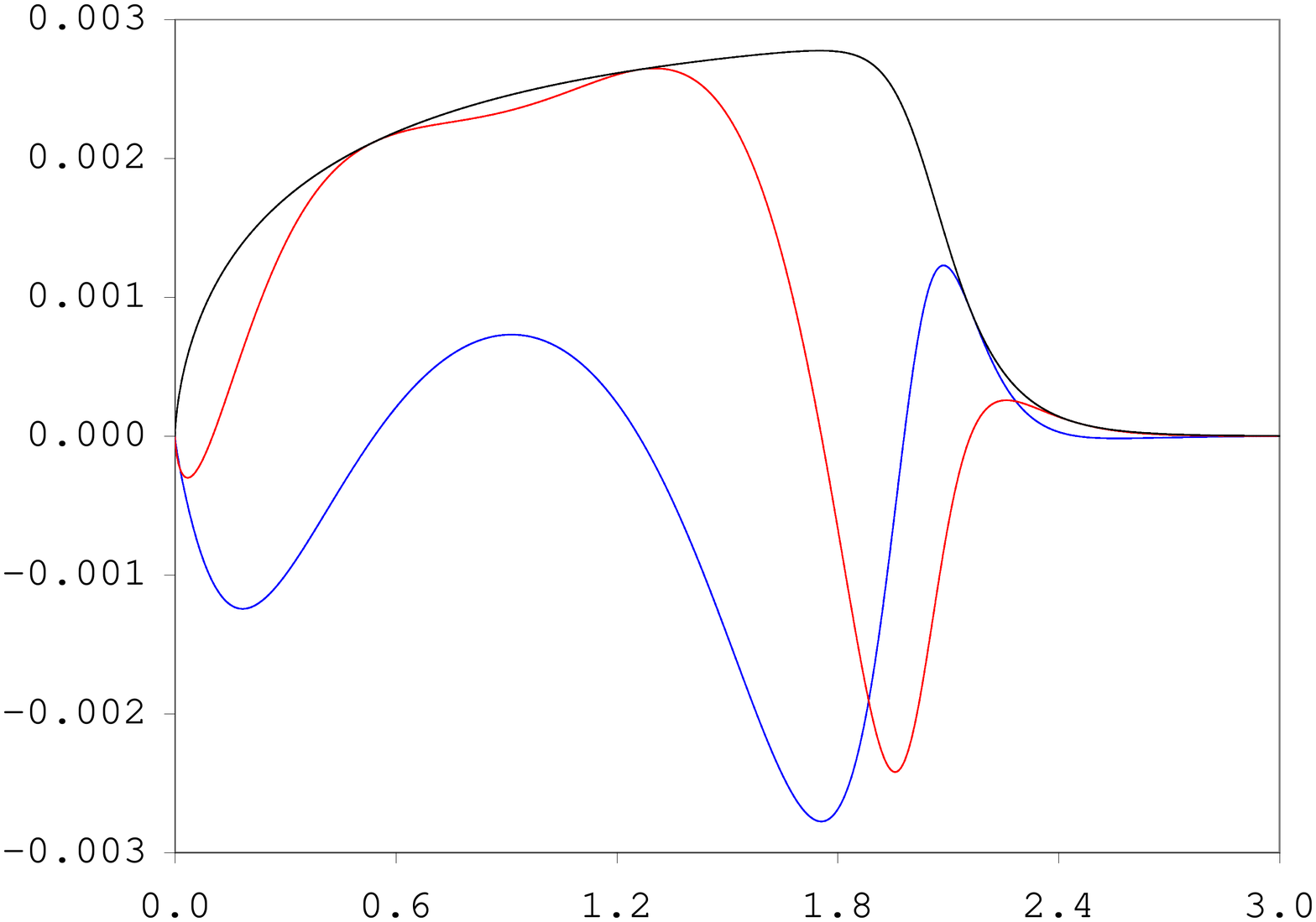}  &  \includegraphics[width=10cm]{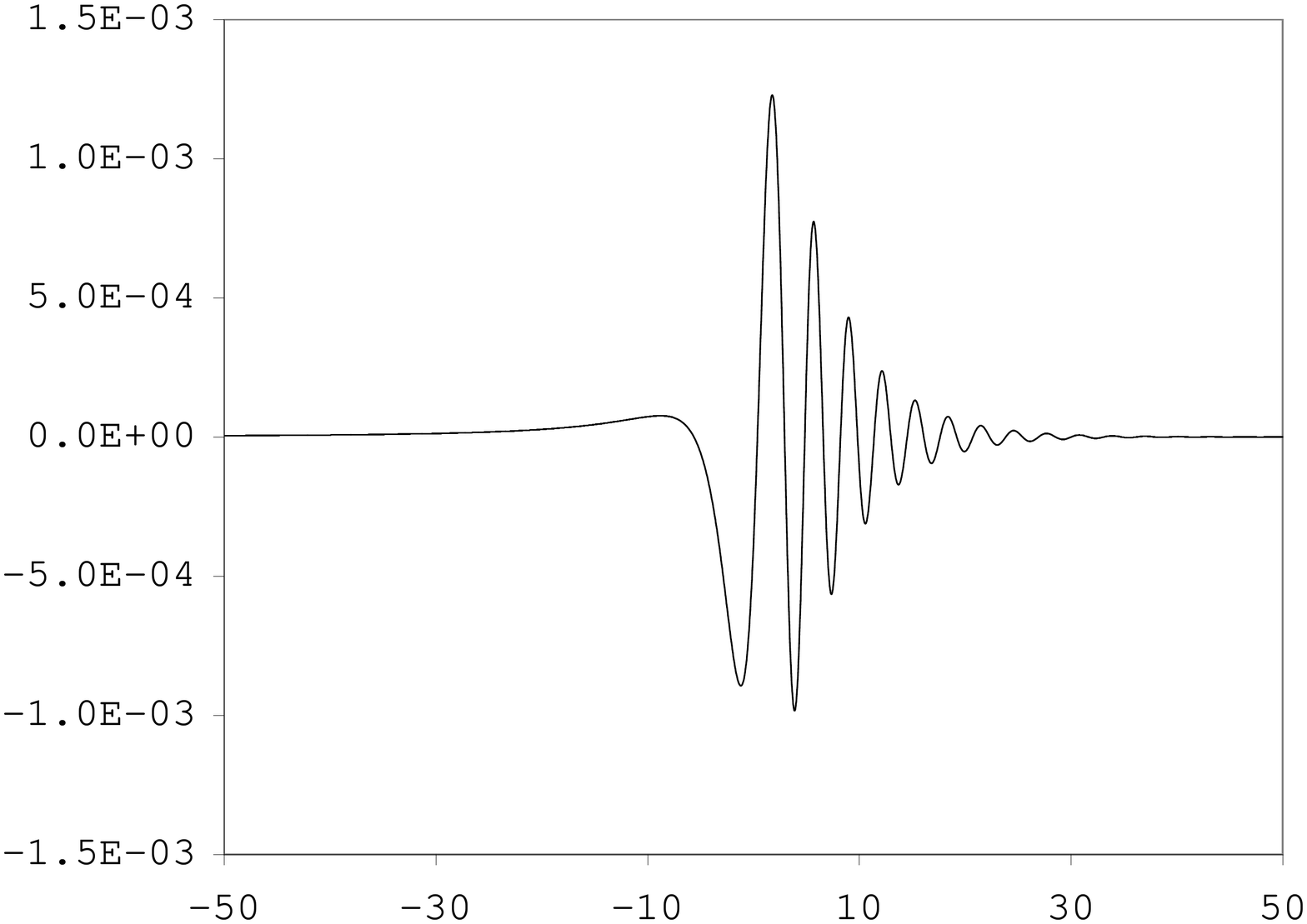} 
\end{tabular}
\vspace{0.5cm}
\centering
\begin{tabular}{c}
$\psi_{5,out}$ QNM fit \\
\includegraphics[width=10cm]{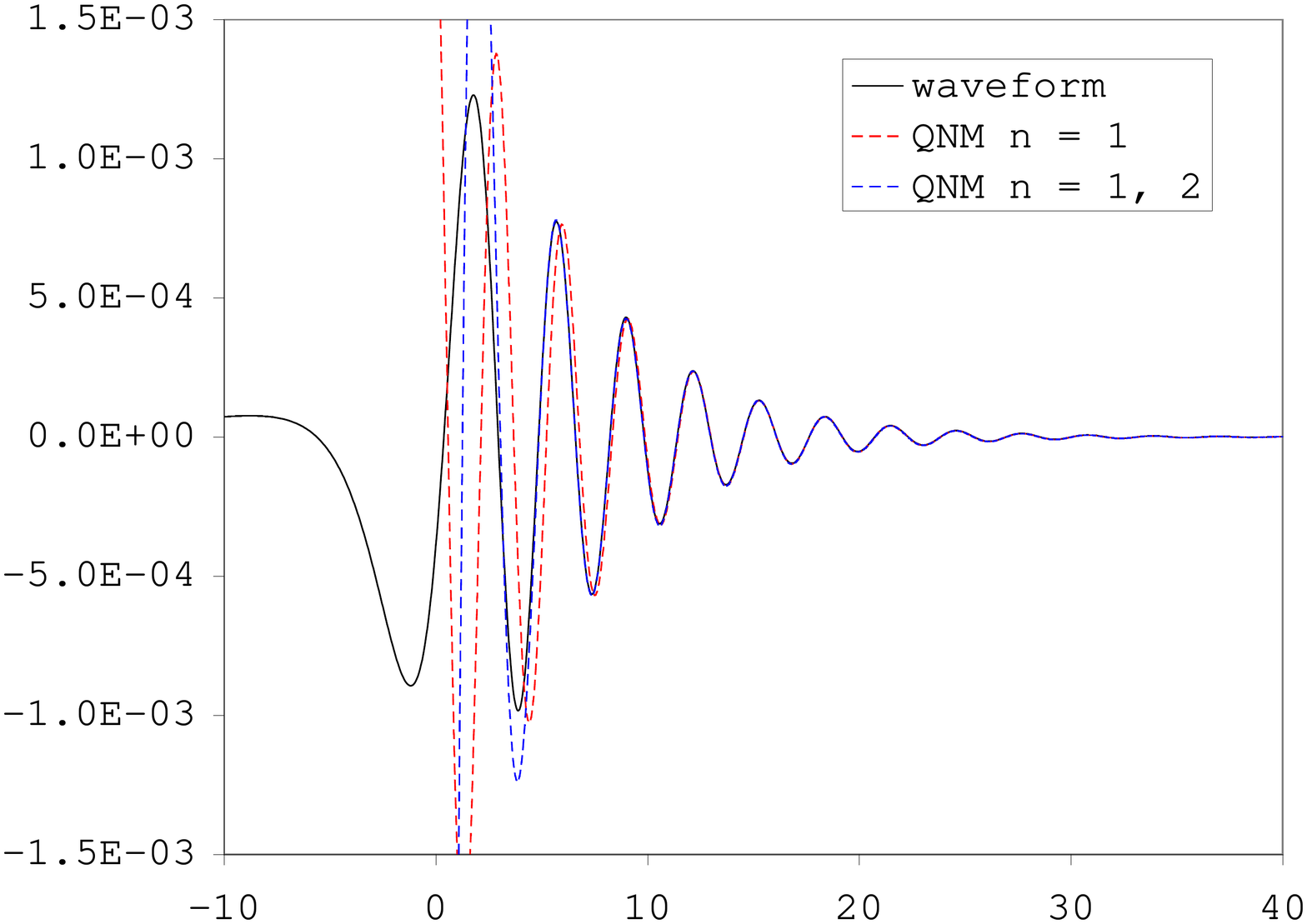}  
\end{tabular}
\caption[Spectrum, waveform and QNM fit for $l = 5$]{Top left panel: the amplitude
$A_{5,out}/Mm$. Top right panel: the waveform $\psi_{5,out}/m$. Bottom panel: the curves
corresponding to the QNM contribution up to the second overtone $n = 2$.}
\label{fig:l5}  
\end{sidewaysfigure}

\begin{sidewaysfigure} 
\begin{tabular}{cc}
$A_{6,out}/Mm$ vs. $2M\omega$  & $\psi_{6,out}/m$ vs. $u/2M$      \\
\includegraphics[width=10cm]{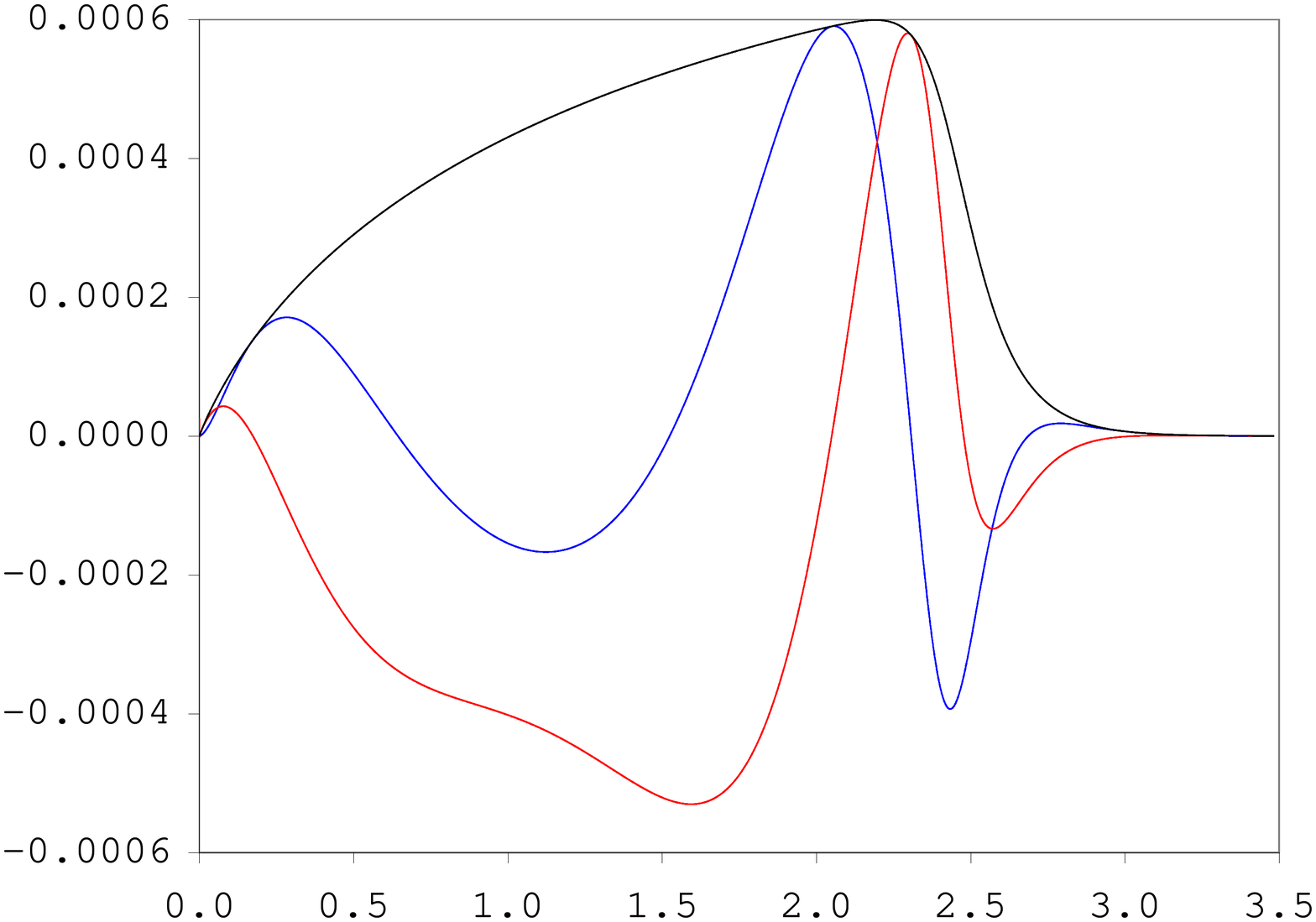}  &  \includegraphics[width=10cm]{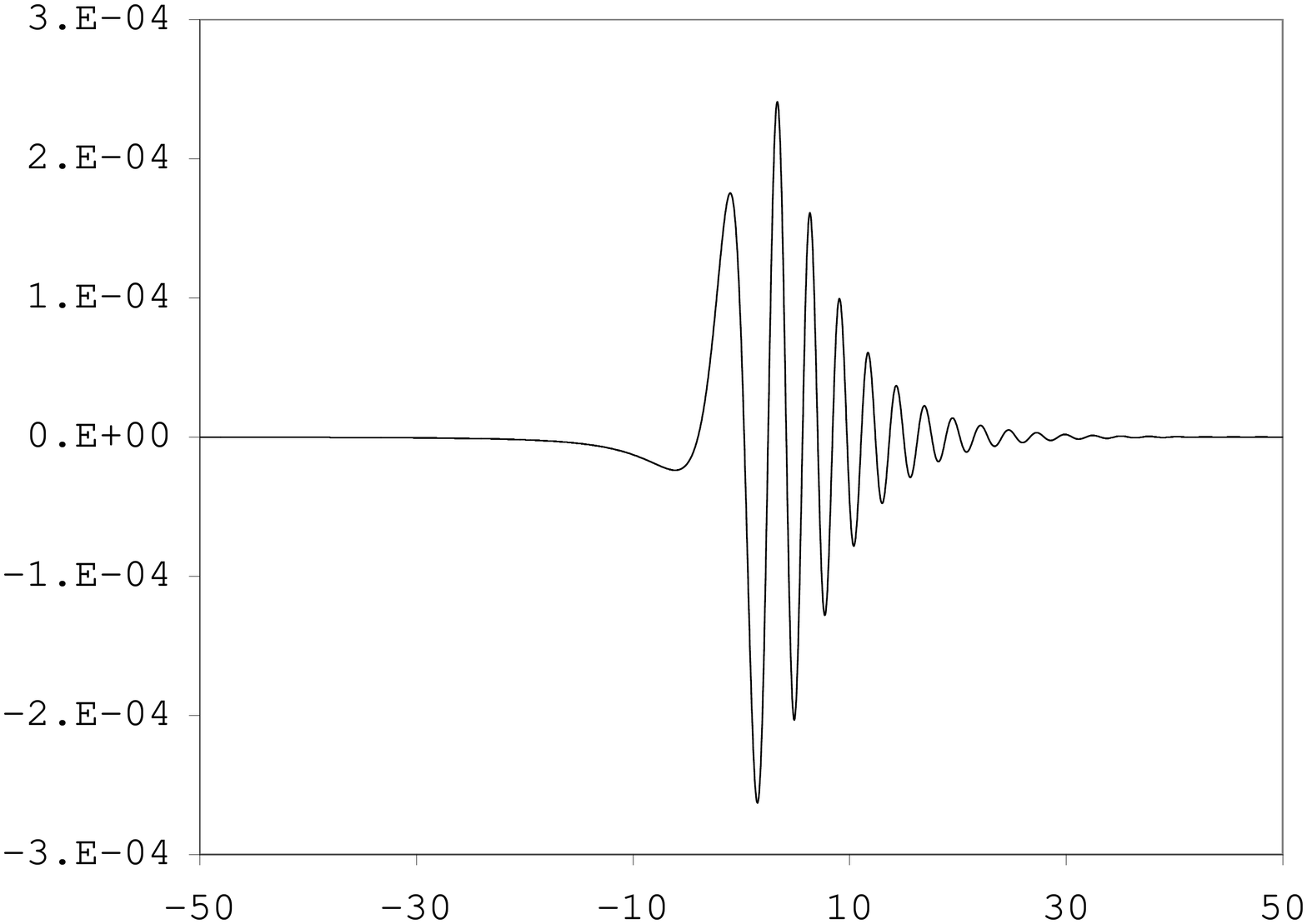} 
\end{tabular}
\vspace{0.5cm}
\centering
\begin{tabular}{c}
$\psi_{6,out}$ QNM fit \\
\includegraphics[width=10cm]{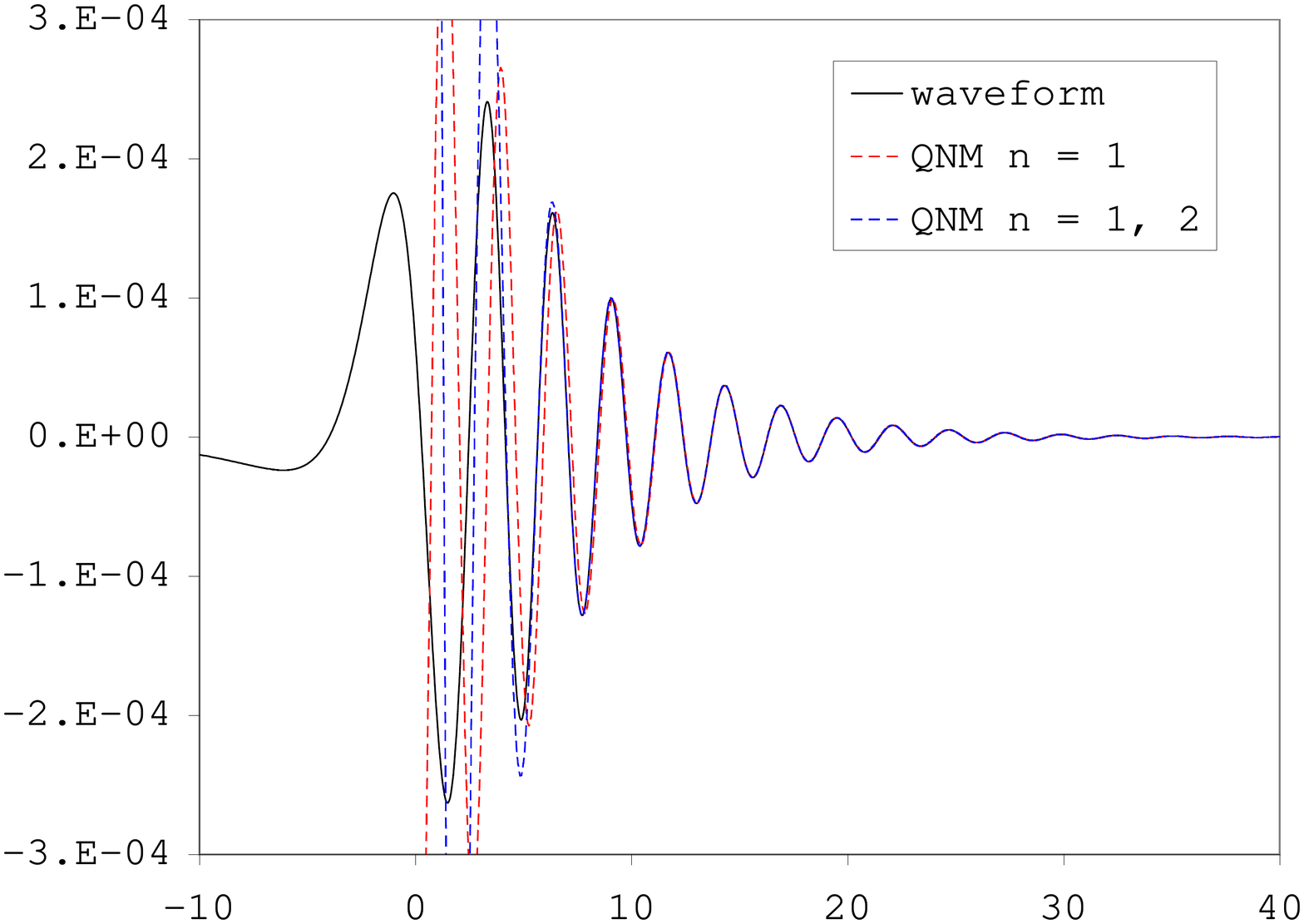}    
\end{tabular}
\caption[Spectrum, waveform and QNM fit for $l = 6$]{Top left panel: the amplitude
$A_{6,out}/Mm$. Top right panel: the waveform $\psi_{6,out}/m$. Bottom panel: the curves
corresponding to the QNM contribution up to the second overtone $n = 2$.}
\label{fig:l6}  
\end{sidewaysfigure}

\begin{figure}
\centering
\begin{tabular}{c}
$2M\omega_{n,\Im}$ vs. $2M\omega_{n,\Re}$  \\
\includegraphics[width=13cm]{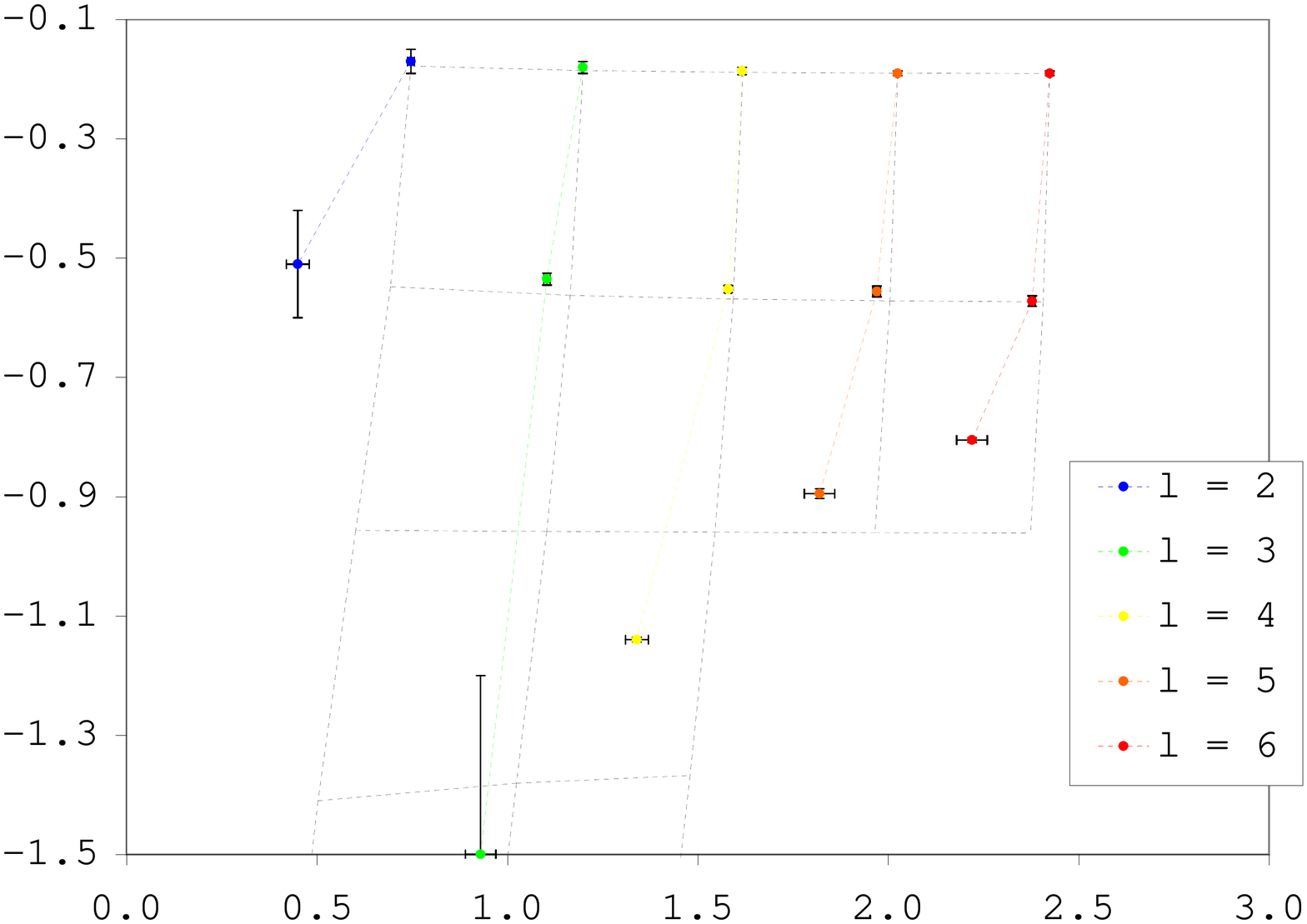} 
\end{tabular}
\begin{tabular}{cc}
$B_{l,n}/Mm$ vs. overtone number $n$  &  $\theta_{l,n}$ vs. overtone number $n$  \\
\includegraphics[width=8cm]{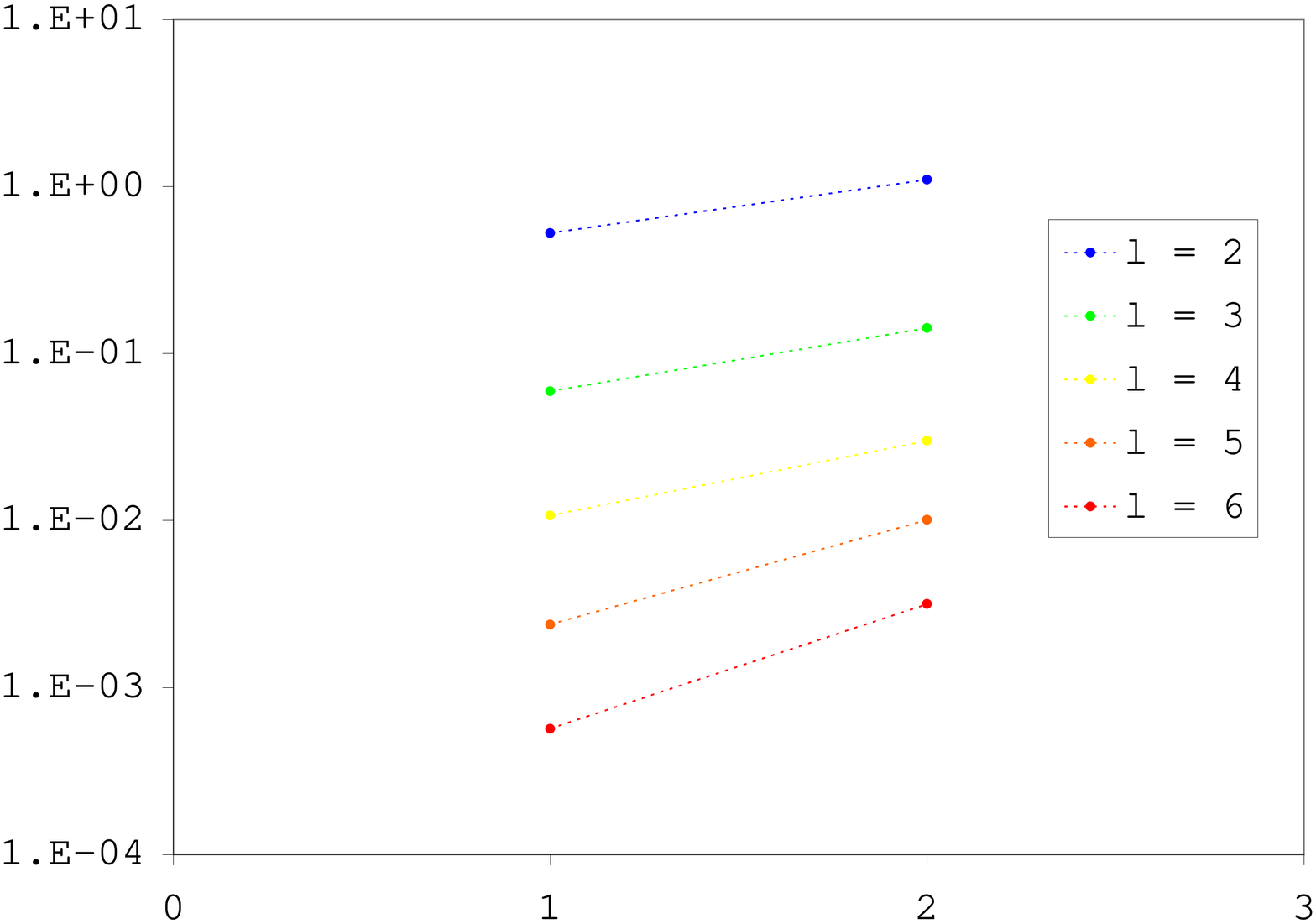}  &  \includegraphics[width=8cm]{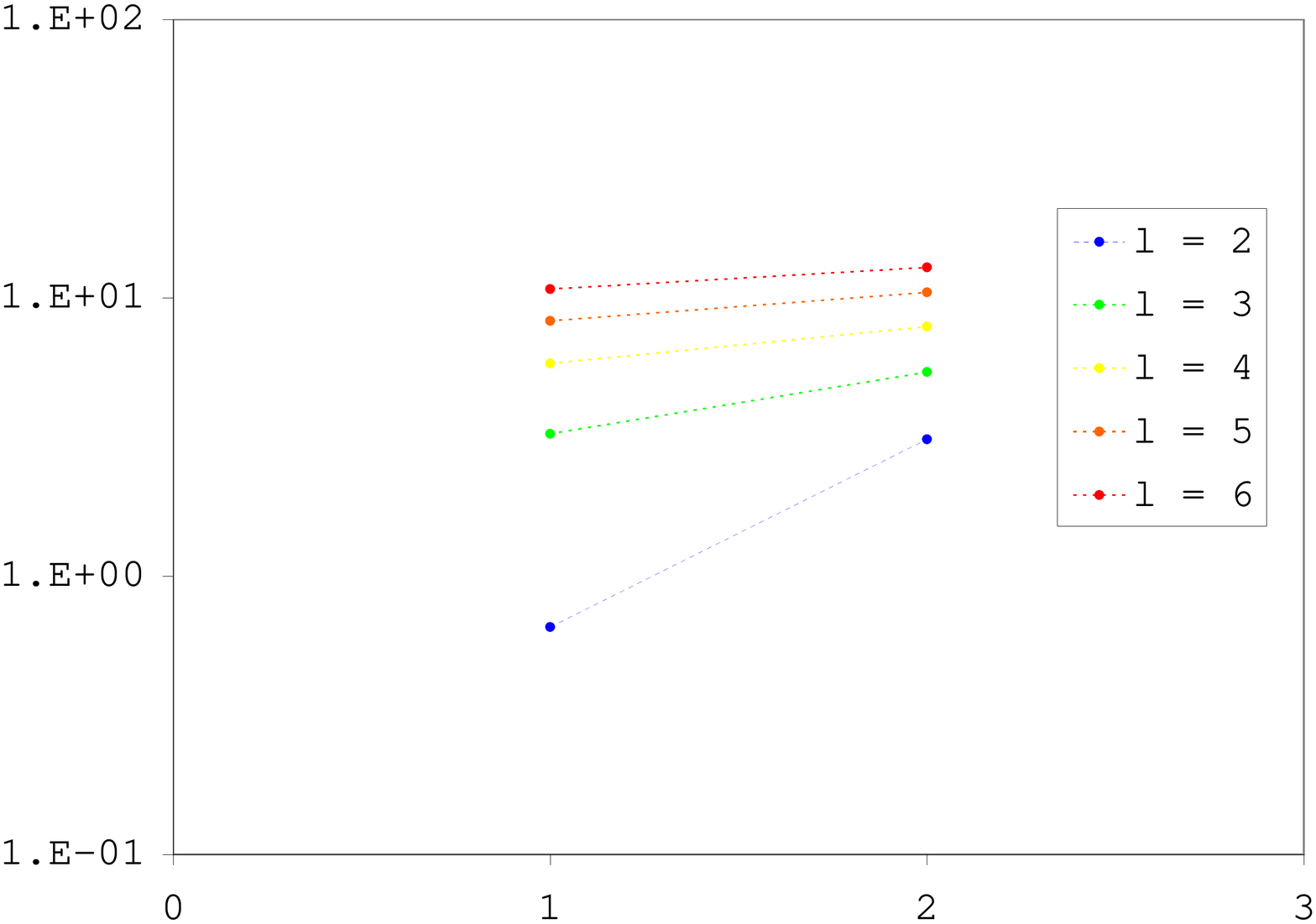} \\
 \\
$\omega_{3,1,\Re}$ vs. $[a,b]$  &  $|\{$ points in the $k$-th slice $\}|$ vs. $k$  \\
\includegraphics[width=8cm]{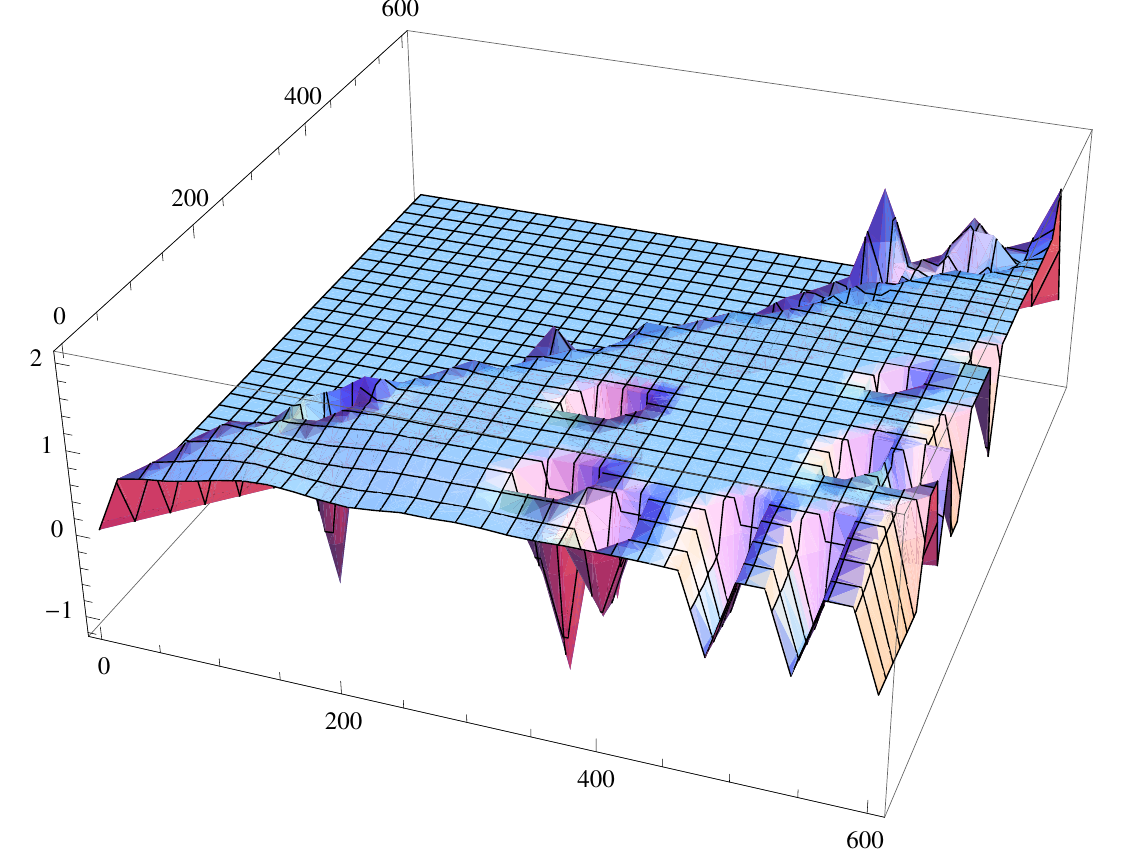}  &  \includegraphics[width=8cm]{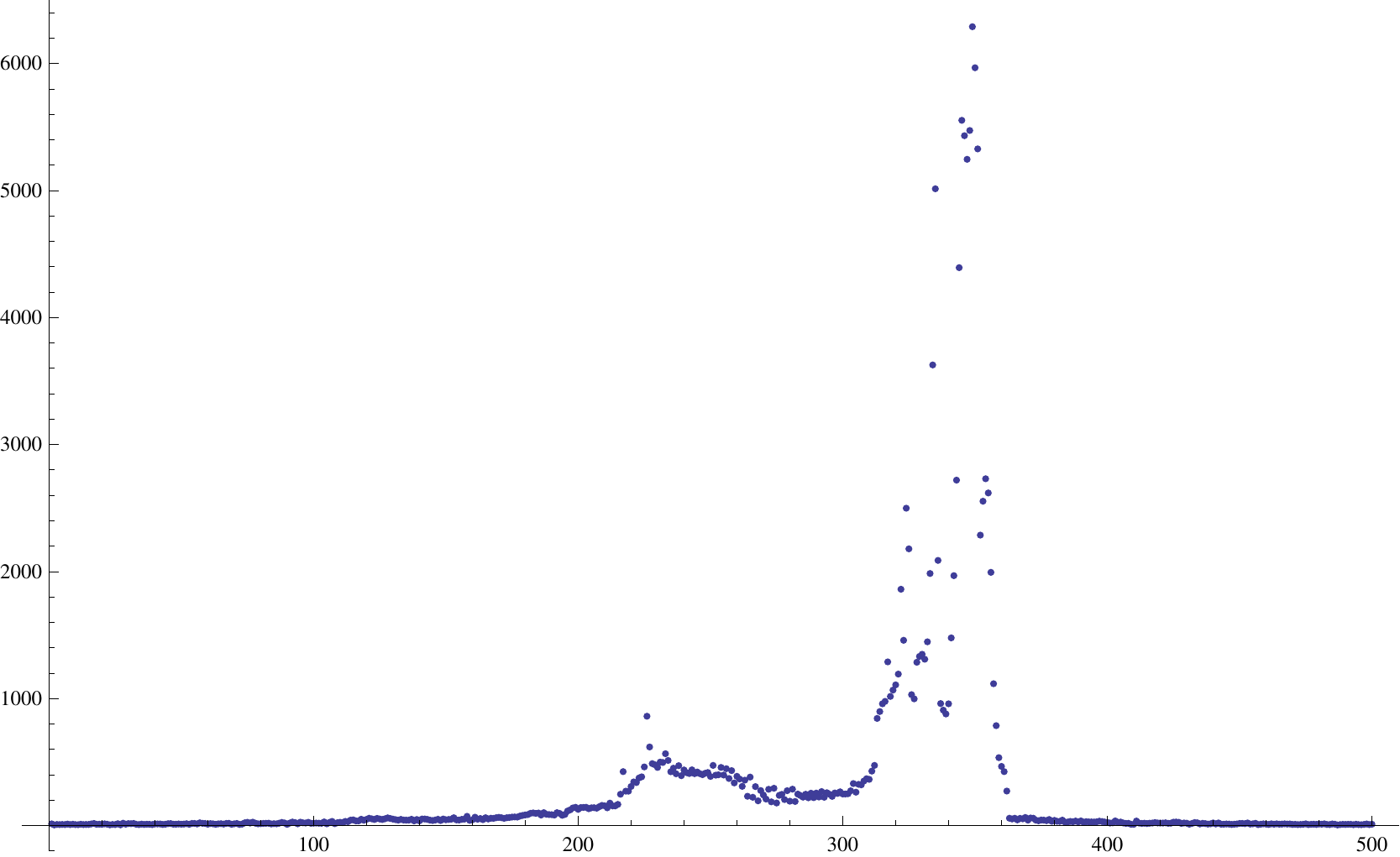} 
\end{tabular}
\caption[QNM related figures]{Top panel: the computed QNMs
through the waveforms fit. The dotted lines' intersections are the expected
values. Vertical lines link QNMs of the same $l$ while horizontal lines link
QNMs with the same $n$. $n$ increases by going downwards. Middle panels: the amplitudes $B_{l,n}/Mm$ and phases $\theta_{l,n}$ of the first two overtones. Bottom left panel: example of the
3D plot obtained when fitting the waveform for QNMs on all intervals $[a,b]$ in
$[0,30]$ of $u/2M$. Bottom right panel: the corresponding histogram giving the number
of points in each horizontal slice of the 3D plot.}
\label{fig:QNM}
\end{figure}

\end{center}

\end{document}